\documentclass[twocolumn]{aastex631}
\shorttitle{Diverse Rotation Curves of Galaxies in a Simulated Universe}
\shortauthors{Jeong et al.}
\usepackage{booktabs}
\usepackage{float}

\begin{document}

\title{Diverse Rotation Curves of Galaxies in a Simulated Universe: the Observed Dependence on Stellar Mass and Morphology Reproduced}

\correspondingauthor{Ho Seong Hwang}
\email{hhwang@astro.snu.ac.kr}

\author{Daeun Jeong}
\affiliation{Astronomy Program, Department of Physics and Astronomy, Seoul National University, 1 Gwanak-ro, Gwanak-gu, Seoul 08826, \\
Republic of Korea}

\author{Ho Seong Hwang}
\affiliation{Astronomy Program, Department of Physics and Astronomy, Seoul National University, 1 Gwanak-ro, Gwanak-gu, Seoul 08826, \\
Republic of Korea}
\affiliation{SNU Astronomy Research Center, Seoul National University, 1 Gwanak-ro, Gwanak-gu, Seoul 08826, Republic of Korea}
\affiliation{Australian Astronomical Optics - Macquarie University, 105 Delhi Road, North Ryde, NSW 2113, Australia}

\author{Haeun Chung}
\affiliation{University of Arizona, Steward Observatory, 933 N Cherry Avenue, Tucson, AZ 85721, USA}

\author{Yongmin Yoon}
\affiliation{Korea Astronomy and Space Science Institute (KASI), 776 Daedeokdae-ro, Yuseong-gu, Daejeon 34055, Republic of Korea}
\affiliation{Department of Astronomy and Atmospheric Sciences, Kyungpook National University, Daegu, 41566, Republic of Korea}



\begin{abstract}

We use the IllustrisTNG cosmological hydrodynamical simulation to study the rotation curves of  galaxies in the local universe. To do that, we first select the galaxies with 9.4 $<$ $\log{(M_\mathrm{star}/M_\odot)}$ $<$ 11.5 to make a sample comparable to that of SDSS/MaNGA observations. We then construct the two-dimensional line-of-sight velocity map and conduct the fit to determine the rotational velocity and the slope of the rotation curve in the outer region ($R_\mathrm{t}<r<3\times r_\mathrm{half,*}$). The outer slopes of the simulated galaxies show diverse patterns that are dependent on morphology and stellar mass. The outer slope increases as galaxies are more disky, and decreases as galaxies are more massive, except for the very massive early-type galaxies. The outer slope of the rotation curves shows a correlation with the dark matter fraction, slightly better than for the gas mass fraction. Our study demonstrates that the observed dependence of galaxy rotation curves on morphology and stellar mass can be successfully reproduced in cosmological simulations, and provides a hint that dark matter plays an important role in shaping the rotation curve. The sample of simulated galaxies in this study could serve as an important testbed for the subsequent study tracing galaxies back in time, enabling a deeper understanding of the physical origin behind the diverse rotation curves.
\end{abstract}

\section{Introduction} \label{sec:intro}

A galaxy rotation curve shows how the rotation velocity of stars or gas in a galaxy varies with the distance from the center of the galaxy. A galaxy rotation curve is an important tool for studying the internal structure of a galaxy (\citealt{Oort40, Sofue&Rubin01}). The square of the amplitude of rotational velocity is proportional to the enclosed mass, so the curve also provides a way to map the mass distribution in a galaxy including invisible dark matter \citep{Rubin70, sofue17}. 

One interesting thing from recent integral field spectroscopic observations of galaxies is that the shape of rotation curves of galaxies is diverse (\citealt{genzel2017strongly, Tiley19, genzel2020rotation}). This diversity might be attributed to dark matter, which constitutes the majority of the mass in the outer regions of galaxies (\citealp{Lovell18})\footnote{It should be noted that there are also other ways to explain galaxy rotation curves without introducing dark matter (e.g. \citealt{milgrom83, yoon23_emergentG})}. In detail, the amount of dark matter and the radial profile of dark matter density can vary depending on different histories of mass accretion in galaxies. It has been observed that high-z spiral galaxies have a decreasing rotation curve that differs from the flat rotation curve of local spiral galaxies (\citealt{genzel2017strongly, Lang_2017, genzel2020rotation}). By determining the cause of this z-dependent rotation curve shape, we can learn how dark matter affects the rotation curve itself and other galaxy properties influencing the rotation curve. This means that unraveling the cause of diverse rotation curves can provide important insights into the role of dark matter in galaxy evolution. 

In this regard, there have been several observational studies focusing on the relation between the shape of rotation curves and the physical properties of galaxies, which include morphological type \citep{Corradi90,Erroz16,Kalinova17} and luminosity (\citealp{casertano91,persic96,sofue01,Noordermeer07,Kalinova17}). Several simulation studies have been conducted to identify the underlying causes of the observed diversity in rotation curve shapes. For example, \citet{oman2015unexpected} and \citet{santos2018nihao} have studied the diverse shapes of rotation curves in the inner region of dwarf galaxies that is related to the cusp-core issue (\citealp{Flores94,Moore94,deBlok08, deBlok10,Oh15}). \citet{oman2015unexpected} show that observed galaxies have more diverse shape at fixed maximum circular velocity than the simulated galaxies. Observed dwarf galaxies which exhibit inner mass deficits, commonly referred to as "cores," do not match well with simulated galaxies. \citet{oman2015unexpected} conclude that this difference could come from several possibilities such as the complexity of the dark matter properties, the limitation of the simulation and observation. \citet{santos2018nihao} compare the galaxies from the NIHAO simulation and the observed galaxies in the SPARC dataset (\citealp{Lelli16}). Due to the dark matter halo expansion in the simulation, the simulated galaxies develop the core, which is often seen in observed galaxies. However, they could not reproduce well the properties of observed starbursts and emission-line galaxies. These starbursts and emission-line galaxies could be intriguing targets offering new insights into the underlying processes. In addition, \citet{ubler2021kinematics} compare kinematics of seven massive galaxies ($M_*>4\times 10^{10}M_\odot$) in TNG50 with that of galaxies observed in \citet{genzel2020rotation}; they analyze the dark matter fraction in galaxies where the center is dominated by baryon, in relatively high-redshift samples (i.e. z$\sim$2). The galaxies in TNG50 exhibit rotational velocities and velocity dispersions that are broadly consistent with observations, as well as low central dark matter fractions. However, at effective radii, TNG galaxies tend to have higher dark matter fractions than the observed galaxies. They conclude that the limited resolution of the simulation makes the difference.

Recently, \citet{yoon2021rotation} use the galaxies in SDSS/MaNGA (\citealp{Wake17}) to conduct a detailed analysis of rotation curves for local galaxies. They investigate the dependence of galaxy rotation curves on both morphological type and stellar mass. They find that the outer part of the galaxy rotation curve increases as galaxies are more disky, and decreases as galaxies are more massive except for the very early-type galaxies (T-type $<$ -2.5). This interesting result of \citet{yoon2021rotation} is the motivation of our study. Although the results presented in \citet{yoon2021rotation} are intriguing, due to limitations in observational data, it is challenging to explore the correlation between various galactic properties and the shape of the rotation curves. This limitation can be overcome with the simulation data in this study, which can provide a more complete and richer interpretation of the rotation curves and its connection to galaxy properties. Therefore, we would like to examine whether we could reproduce the trends in observations from cosmological simulations, and to understand the physical origin behind the diversity of galaxy rotation. This is the first study of our group in this direction. 



Throughout this study, we adopt the $\Lambda$ cold dark matter ($\Lambda$CDM) cosmological model with $H_0 = 67.74$ km s$^{-1}$Mpc$^{-1}$, $\Omega_\Lambda$= 0.6911, and $\Omega_m$= 0.3089 as in IllustrisTNG (\citealp{nelson2021illustristng, Plank2016}).

\section{Data and Analysis} \label{sec:data&analysis}
\subsection{The TNG100 simulation} \label{subsec:TNG}

The llustrisTNG project (\citealp{Pillepich_2017,Nelson_2017,Marinacci_2018,Naiman_2018,Springel_2017}) comprises a suite of large-volume, cosmological, gravo-magnetohydrodynamical simulations conducted with the moving-mesh code \textit{Arepo} (\citealp{springel2010moving}). Each simulation self-consistently solves for the coupled evolution of dark matter, cosmic gas, luminous stars, and supermassive black holes from early cosmic epochs to the present day (z = 0) using a comprehensive model for galaxy formation physics (\citealp{nelson2021illustristng}). The TNG suite includes three flagship simulations, namely TNG50, TNG100, and TNG300, which are accompanied by a rich set of halo/subhalo catalogs, merger trees, and dark-matter-only counterparts. 

Among the three versions of TNG simulations, we select TNG100 due to its relatively large-volume and high-resolution, which are crucial for making a large galaxy sample with accurate measurement of galaxy rotation curves. In the case of TNG50, the smaller volume results in a small number of galaxies, which limits its utility for statistical analyses. Conversely, TNG300 suffers from insufficient mass resolution, making it difficult to obtain reliable velocity maps for low-mass galaxies. The TNG100 has a large cubic volume of (106.5 Mpc)$^3$ that is big enough to investigate a large sample of galaxies. Among the TNG100 simulation runs, the highest resolution realization, TNG100-1, includes 2$\times 1820^3$ resolution elements. The dark matter and baryonic mass resolutions are $7.5\times 10^6 M_\odot$ and $1.4\times 10^6 M_\odot$, respectively. The gravitational softening length for dark matter and stars is 0.74 kpc at z=0. 

\subsection{Sample selection} \label{subsec:sample}
To investigate the properties of low-redshift galaxies as for the MaNGA program, we choose the Snapshot99 simulation data which corresponds to the zero redshift (i.e. z=0.0)\footnote{The median redshift of the MaNGA galaxy sample in Yoon et al. (2021) is 0.037.}. We use only the galaxies of cosmological origin: i.e. $SubhaloFlag$=1. To select the simulated galaxies that are comparable to the observational sample, we choose the galaxies within the stellar mass range of 9.4 $< \log{(M_{*}/M_\odot)} <$ 11.5 as in \citet{yoon2021rotation}. The stellar mass ($M_\mathrm{star}$) of a galaxy is determined by summing the masses of the stellar particles of the subhalo within twice the stellar half-mass radius (2 $r_\mathrm{half,*}$). The steller half-mass radius ($r_\mathrm{half,*}$) is defined as the radius containing half of the total stellar mass of the subhalo (\citealp{Pillepich_2017}). At z=0, a total of 12,009 galaxies within the TNG100 volume satisfy the criteria.

\subsection{Construction of 2D Velocity Map} \label{subsec:Vmap}
\begin{figure}[b]
\centering
\includegraphics[width=80mm,scale=1]{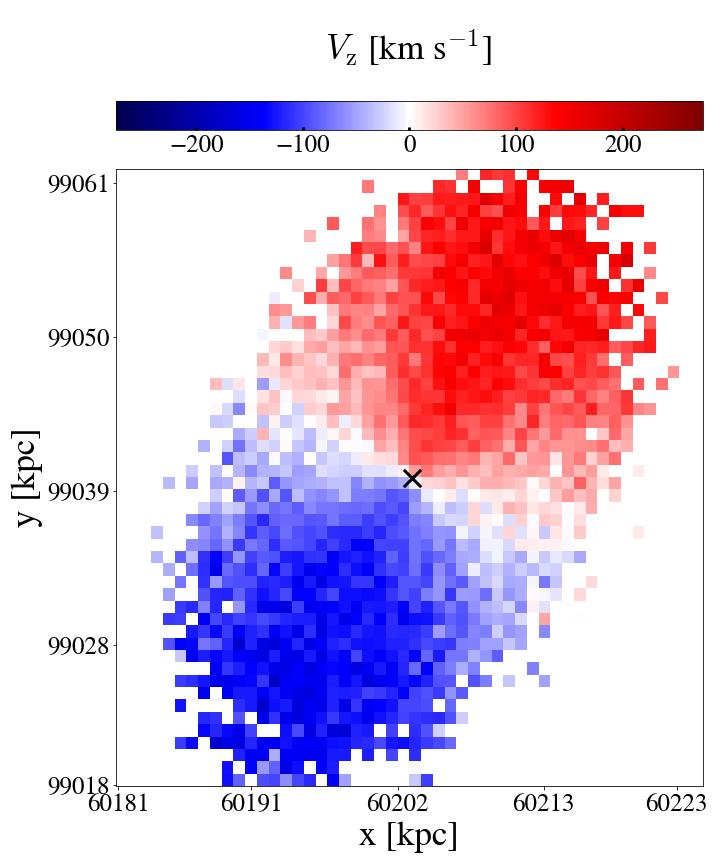}
\caption{Constructed 2D line-of-sight velocity map of a subhalo (Subhalo ID = 88673). The line-of-sight direction is Z-axis of the simulation box. The X- and Y-axis denote the x and y coordinates of the simulation. The color bar represents the line-of-sight velocity ($V_\mathrm{z}$) and ranges from -1.5 $V_c$ to +1.5 $V_c$. The systematic line-of-sight velocity at the kinematic center is set to 0 km s$^{-1}$. The cross symbol ($\times$) in black color shows the spatial position of the kinematic center of the subhalo.   \label{fig:Vmap}}
\end{figure}
We create mock velocity maps of each galaxy for three different lines of sight. One of the lines of sight corresponds to the Z-axis direction of the simulation box, where the galaxy is observed on the x-y plane. The remaining lines of sight are aligned with the X-axis and Y-axis directions of the simulation. 

For each line of sight, we bin the velocity map to resemble the MaNGA datacube which is used for rotation curve fitting from \citet{yoon2021rotation}. The size of the velocity map is set to ensure that the length along the major axis of the galaxy is equal to 3 times the stellar half-mass radius (3 $r_\mathrm{half,*}$) measured from the galactic center. This is to roughly match the spatial coverage of the MaNGA observational data; the MaNGA has radial coverage of 1.5 $R_e$ for two-thirds of the final sample or 2.5 $R_e$ for one-third of the final sample. The stellar half-mass radius ($r_\mathrm{half,*}$) of a galaxy is comparable to half-light radius ($R_e$) of the galaxy over the range of $10^8 < M_\mathrm{star}/M_\odot < 10^{11.5}$ (\citealp{schaye2015eagle}). In the result, the radial coverage in this study is slightly larger than that of MaNGA observations, but should not have significant impact on our main conclusions as long as the outer part of the rotation curve changes linearly (to be discussed in Section \ref{subsec:observ}).


The datacube sampling of the MaNGA is 0.5$"\times$0.5$"$. Considering the median redshift of the MaNGA sample in \citet{yoon2021rotation}, this gives the physical spaxel size of 0.367 kpc at the distance of z=0.037. To match the physical spaxel size with observation, we bin the velocity map into a 50 $\times$ 50 grid\footnote{The median value of stellar half-mass radius for the TNG galaxies is 3.53 kpc.}. We then take the median of the line-of-sight velocities ($V_\mathrm{LoS}$) of the star particles in each pixel to make the mock velocity map. 

Figure \ref{fig:Vmap} shows an example of the constructed 2D velocity map. The line-of-sight is parallel to the Z-axis of the simulation. The color bar represents the line-of-sight velocity $V_\mathrm{z}$, which is set to 0 km s$^{-1}$ for the systematic velocity of the galaxy. The red and blue color denote receding and approaching velocity, respectively. Considering the Poisson noise, we include only the pixels that contain more than 9 star particles in them for the fit (i.e. S/N$\geq$3). Because of this signal-to-noise ratio criterion, the effective radial coverage of the pixels containing reliable velocity information is sometimes smaller than three times the stellar half-mass radius that gives the size of the velocity map. We will discuss this effect in Section \ref{subsec:circ} in detail.

\subsection{Rotation Curve Model Fitting} \label{subsec:RCfitting}

In this section, we outline the methodology employed to fit 2D rotation curve (RC) models to the line-of-sight velocity maps constructed in Section \ref{subsec:Vmap}. We use an empirical RC model function which can fit various kinds of rotation curves. Previous studies use various functional forms including arctangent, exponential, and hyperbolic tangent to describe rotation curves that rise from the center and remain constant at larger radius (\citealp{IMAGES, Feng2011, Andersen2013, Bouche2015}). However, multiple studies have revealed that the outer slope of RCs varies depending on the properties of the galaxies (\citealp{Corradi90, casertano91, persic96, sofue01, Noordermeer07, Kalinova17}). Therefore, we adopt the functional form that is the same as the one in \citet{yoon2021rotation} and \citet{Chung_2021}, which can effectively capture both the inner rising part and the variations in the outer region of the RC. The chosen functional form is a combination of the hyperbolic tangent function and a linear term as follows:
\begin{equation}
    V(r) = V_c \tanh{(r/R_\mathrm{t})} + s_\mathrm{out}r  \label{eq:Vr}
\end{equation}

\begin{figure}[t!]
\centering
\includegraphics[width=85mm,scale=1]{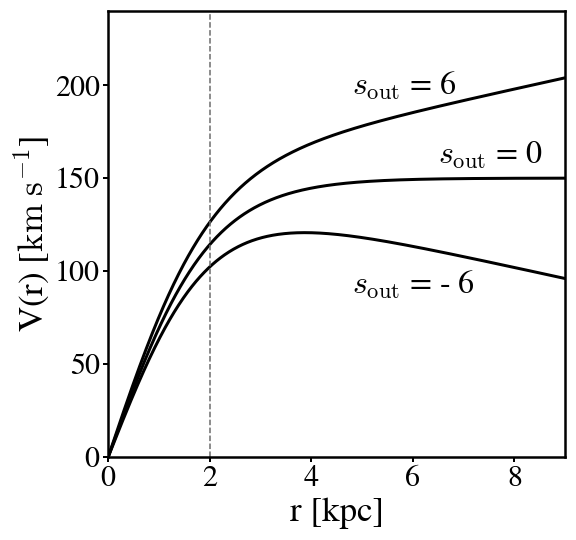}
\caption{Examples of three rotation curves in the form of Equation (\ref{eq:Vr}). All of the rotation curves have $V_c$ = 150 km s$^{-1}$ and $R_\mathrm{t}$ = 2 kpc but have different $s_\mathrm{out}$ (6, 0, and -6 km s$^{-1}$ kpc$^{-1}$, respectively). The vertical dashed line denotes $R_\mathrm{t}$ = 2 kpc. We note that the rotation curve with $s_\mathrm{out}$ = -6 km s$^{-1}$ kpc$^{-1}$ does not reach 150 km s$^{-1}$ due to the negative linear term in Equation (\ref{eq:Vr}), even though $V_c$ is 150 km s$^{-1}$. (\citealp{yoon2021rotation})   \label{fig:RCsoutex}}
\end{figure}

Here, $s_\mathrm{out}$ represents the slope of the RC at large radii r $\gg R_\mathrm{t}$. $R_\mathrm{t}$ denotes a turnover radius, where the hyperbolic tangent term switches to a flat region. The coefficient $V_c$ in the hyperbolic tangent term equals to the maximum V(r) when $s_\mathrm{out}$ is 0. Figure \ref{fig:RCsoutex} depicts three examples of RCs generated from Equation (\ref{eq:Vr}) with varying values of $s_\mathrm{out}$.

To fit our RC model to the 2D line-of-sight velocity maps constructed in Section \ref{subsec:Vmap}, we employ a two-dimensional model based on previous works (\citealp{begeman1989hi,beckman04,oh18,Chung_2021}). The specific form of the model is as follows:

\begin{equation}
    V_\mathrm{obs}(r',\phi') = V_\mathrm{sys} + V(r) \sin {i} \cos{(\phi -\phi_0)}  \label{eq:Vobs}
\end{equation}

Here, V($r'$, $\phi'$) represents the projected 2D line-of-sight velocity map, where $r'$ denotes the radial distance from the kinematic center of a galaxy to each pixel on the plane of the sky, and $\phi'$ represents the position angle of the pixel on the plane of the sky. The parameters r and $\phi$ correspond to the radial coordinate and position angle in the plane of the galaxy (the deprojected plane), respectively. The position angle refers to the angle of the pixel measured counter-clockwise from the vertical axis of the image. The function V(r) corresponds to the chosen RC function in Equation (\ref{eq:Vr}). The parameter $i$ represents the kinematic inclination angle, while $\phi_0$ denotes the kinematic position angle of the galaxy. Lastly, $V_\mathrm{sys}$ corresponds to the systematic line-of-sight velocity at the kinematic center. The fitting process involves the utilization of eight parameters, namely $x_\mathrm{center}$, $y_\mathrm{center}$, $V_\mathrm{sys}$, $i$, $\phi_0$, $V_c$, $R_\mathrm{t}$, and $s_\mathrm{out}$. These parameters encompass the position of kinematic center ($x_\mathrm{center}$ and $y_\mathrm{center}$) in the plane of the sky, $V_\mathrm{sys}$, $i$, $\phi_0$, and the parameters $V_c$, $R_\mathrm{t}$, and $s_\mathrm{out}$ from the RC model.

We perform the non-linear least square fit to the two-dimensional velocity map with the model (i.e. equations \ref{eq:Vr} and \ref{eq:Vobs}) using the curve$\_$fit function in $scipy$ package (\citealp{SciPy}). We use only the pixels that contain star particles more than 9 for the fit as mentioned in Section \ref{subsec:Vmap}.

To provide reasonable initial values for the fitting parameters, we use the information on subhaloes (i.e. corresponding to galaxies in observations) provided by the TNG team. For example, the $SubhaloVel$ field is used for the initial velocity of $V_\mathrm{sys}$ parameter. For $x_\mathrm{center}$ and $y_\mathrm{center}$,  we use $SubhaloPos$ field that is the spatial position of the particle with the minimum gravitational potential energy. We use $pafit$ package (\citealp{pafit}) in Python to obtain the initial value of the kinematic position angle of the galaxy. This package gives reliable results as long as the given velocity map is not messy. In the case of inclination angle of the galaxy, we use the ratio between major (a) and minor axis (b) of the galaxy. If a galaxy has a circular disk, the inclination of the disk can be described as $\cos{i}=b/a$. To derive the major and minor axis of galaxy, we determine the galaxy shape using the moments of the spatial distribution of star particles (\citealp{trumpler1953statistical, carter1980morphology, burgett2004substructure, hwang2007searching}). 

Because we conduct the fit along three independent lines of sight, there are three separate sets of fitted results for each galaxy. Among the total 12,009 galaxies, 11,554, 11,567, and 11,544 galaxies have yielded converged fitting results for $V_\mathrm{x}$, $V_\mathrm{y}$, and $V_\mathrm{z}$ lines of sight, respectively. The cases where the fit fails to converge are mainly because of small number of pixels in the velocity map, making it difficult to achieve a satisfactory fit. 

\subsection{Construction of Galaxy Sample with Reliable Fits}
\label{subsec:reliablefits}
There are many cases where the fit looks strange even though the fit converges. We could try different setups to make the fit converge, but decide to exclude the following cases where we could not fix it in the end. 

The first case is the fit that converges with a wrong position angle. To remove such cases, we compute the difference between the fitted value of the kinematic position angle and the initial value obtained from the $pafit$ package (\citealp{pafit}). We confirm that the initial value obtained from the $pafit$ package is accurate from the visual inspection of the velocity map. Therefore, we discard the galaxies when the difference is larger than 20 degrees from the histogram of the difference (not shown in this paper). Applying this condition, the number of galaxies is 10,221, 10,239, and 10,220 for $V_\mathrm{x}$, $V_\mathrm{y}$, and $V_\mathrm{z}$ lines of sight, respectively.

We also eliminate the cases where the fitted parameters reach the boundary values of $R_\mathrm{t}$ and $V_c$. To ensure physically reasonable fittings, we impose bounds on the $R_\mathrm{t}$ and $V_c$ parameters. The range for $R_\mathrm{t}$ is set from 0 to 50 kpc, while the range for $V_c$ is set from 0 to 1500 km s$^{-1}$. These bounds are chosen generously based on known physical constraints and reasonable expectations for the parameter values. The imposed bounds serve as a means to identify and exclude the results that fall outside the expected physical range. After removing these samples, the number of galaxies is 8341, 8425, and 8337 for $V_\mathrm{x}$, $V_\mathrm{y}$, and $V_\mathrm{z}$ lines of sight, respectively.

\citet{Chung_2021} discuss the result when the 2D velocity data do not have sufficient radial coverage to constrain the RC slope at the outer radius. They conduct the tests using the simulated Integral Field Unit (IFU) data, and determine that the maximum radial distance of the pixels in the velocity map should be greater than 2.5 times $R_\mathrm{t}$ to obtain a reliable $s_\mathrm{out}$ value. Therefore, we select the galaxies satisfying $R_\mathrm{max}/R_\mathrm{t}>2.5$. Additionally, we impose a condition that $R_\mathrm{t}$ should be larger than the pixel size of the corresponding velocity map. The fitted value of the radius smaller than the pixel size in the velocity map is unreliable, as the pixel size represents a spatial scale within which the velocity information is captured. Using these conditions, the galaxies with insufficient coverage in velocity maps are removed and the number of remaining galaxies for the $V_\mathrm{x}$, $V_\mathrm{y}$, and $V_\mathrm{z}$ line-of-sight velocities is 5493, 5537, and 5423, respectively. 

The rotation curves for the galaxies with low kinematic inclinations which are close to face-on can exhibit higher uncertainties in the derived parameters. This is primarily attributed to the strong coupling between $\sin{i}$ and $V_c$ terms in Equations (\ref{eq:Vr}) and (\ref{eq:Vobs}), as discussed in previous studies (\citealp{Noordermeer07,Chung_2021}). We also exclude the galaxies with high inclination which are edge-on galaxies due to bad fitting qualities. While \citet{Chung_2021} demonstrate that reliable determination of RCs is feasible for kinematic inclinations $i \geq 20^\circ\text{-}25^\circ$, we expand our selection criteria to include more galaxies with $ 10^\circ < i < 80^\circ$, which is confirmed by visual inspection of the fit. With the inclusion of these criteria, the number of galaxies for the $V_\mathrm{x}$, $V_\mathrm{y}$, and $V_\mathrm{z}$ line-of-sight velocities is 5006, 5073, and 4939, respectively.

We further select the galaxies using the condition that the ratio between the fitted value of two parameters (i.e. $R_\mathrm{t}$ and $V_c$) and its error should be larger than three; this condition is introduced during the visual inspection of the fitting result. The resulting number of galaxies is reduced to 3066, 3099, and 2993 for $V_\mathrm{x}$, $V_\mathrm{y}$, and $V_\mathrm{z}$ results, respectively.

\begin{table}
\centering
\caption{The number of galaxies for different lines of sight ($V_\mathrm{x}$, $V_\mathrm{y}$, $V_\mathrm{z}$) when each selection criterion is applied in order. See Section \ref{subsec:sample} and \ref{subsec:reliablefits} for further details.}
\label{table:sample}
\begin{tabular}{|c|c|c|c|}
\hline
Selection criterion & $V_\mathrm{x}$ & $V_\mathrm{y}$ & $V_\mathrm{z}$ \\
\hline
\hline
Stellar mass cut & 12,009 & 12,009 & 12,009\\
fit converged & 11,554 & 11,567 & 11,544 \\
$\phi_0-\phi_\mathrm{0,init}<20 ^{\circ}$ & 10,221 & 10,239 & 10,220 \\
$V_c$, $R_\mathrm{t}$ bound & 8,341 & 8,425 & 8,337 \\
$R_\mathrm{max}/R_\mathrm{t}>2.5$ & 6,738 & 6,779 & 6,698 \\
$R_\mathrm{t}<$pixel size & 5,493 & 5,537 & 5,423\\
$10^{\circ}<i<80^{\circ}$ & 5,006 & 5,073 & 4,939 \\
$\frac{R_\mathrm{t}}{\mathrm{Err}(R_\mathrm{t})}>3, \frac{V_c}{\mathrm{Err}(V_c)}>3$ & 3,066 & 3,099 & 2,993 \\
$A<0.6$ & 3,033 & 3,072 & 2,971 \\
$P_\mathrm{CR}>0.4$& 3,027 & 3,066 & 2,965 \\
\hline
\end{tabular}
\end{table}
The model we employ, as described in Equations (\ref{eq:Vr}) and (\ref{eq:Vobs}), does not account for non-circular motions in a galaxy. If a galaxy has undergone or is experiencing a merging process, it often displays asymmetric non-circular motions or counter-rotating features, which can be referred to as merging features. Although some galaxies with merging features are already excluded by other conditions before, we further remove the remaining galaxies using two additional criteria. 

First, we exclude the galaxies that show an asymmetry in velocity maps. To do that, we use asymmetry parameter A, which is often derived from galaxy images through the normalization of the difference between the intensity of the original image and the intensity of the image rotated by 180 degrees (\citealp{Schade_1995, Conselice_2000}). Borrowing this concept, we calculate the asymmetry of the velocity maps of our galaxies as follows:
\begin{equation}
    A = \frac{\Sigma | V_\mathrm{LoS,0}+V_\mathrm{LoS,180^\circ}|}{2\Sigma V_\mathrm{LoS,0}}
\end{equation}
After setting the line-of-sight velocity of kinematic center to 0 km s$^{-1}$, $V_\mathrm{LoS,0}$ is the line-of-sight velocity of each pixel in the velocity map, and $V_\mathrm{LoS,180^\circ}$ is its counterpart after a $180^\circ$ rotation of the velocity map. We perform an addition operation instead of subtraction because the line-of-sight velocity can have both positive and negative values unlike the pixel values of galaxy images that are typically non-negative. We decide to exclude the galaxies with asymmetry parameter larger than 0.6 (i.e. A$>$ 0.6), which is determined from the visual inspection of the maps. 

Second, to filter out counter-rotating galaxies from the final sample, we introduce a parameter, $P_\mathrm{CR}$, which represents the probability of a velocity map exhibiting clear rotation. $P_\mathrm{CR}$ is defined as

\begin{equation}
    P_\mathrm{Clear Rotation} = \frac{|N_{ \phi_0 \pm 90^\circ}(V>0)-N_{\phi_0 \pm 90^\circ}(V<0)|}{N_{\phi_0 \pm 90^\circ}}
\end{equation}
 where $N_{\phi_0 \pm 90^\circ}$ is the number of the pixels whose position angle has a difference with $\phi_0$ less than 90 degrees on the velocity map. When the velocity of the kinematic center is set to 0 km s$^{-1}$, $N_{ \phi_0 \pm 90^\circ}(V>0)$ and $N_{\phi_0 \pm 90^\circ}(V<0)$ refer to the number of pixels that have a positive and negative velocity, respectively, satisfying the same position angle condition earlier. We then exclude the galaxies with $P_\mathrm{CR} > 0.4$, which could exhibit counter-rotation or have disordered rotation.

Applying these two criteria, we exclude the galaxies with merging features. After all the aforementioned conditions, we obtain the final samples of 3027, 3066, and 2965 galaxies for the $V_\mathrm{x}$, $V_\mathrm{y}$, and $V_\mathrm{z}$ line-of-sight velocities, respectively. These final samples show satisfactory fit quality with clear rotation, ensuring the reliability of the obtained parameters that characterize the shape of the rotation curve. The number of galaxies that remain after the application of the aforementioned selection criteria is summarized in Table \ref{table:sample}.

\begin{figure*}[hbt!]
\centering
\includegraphics[width=150mm,scale=1]{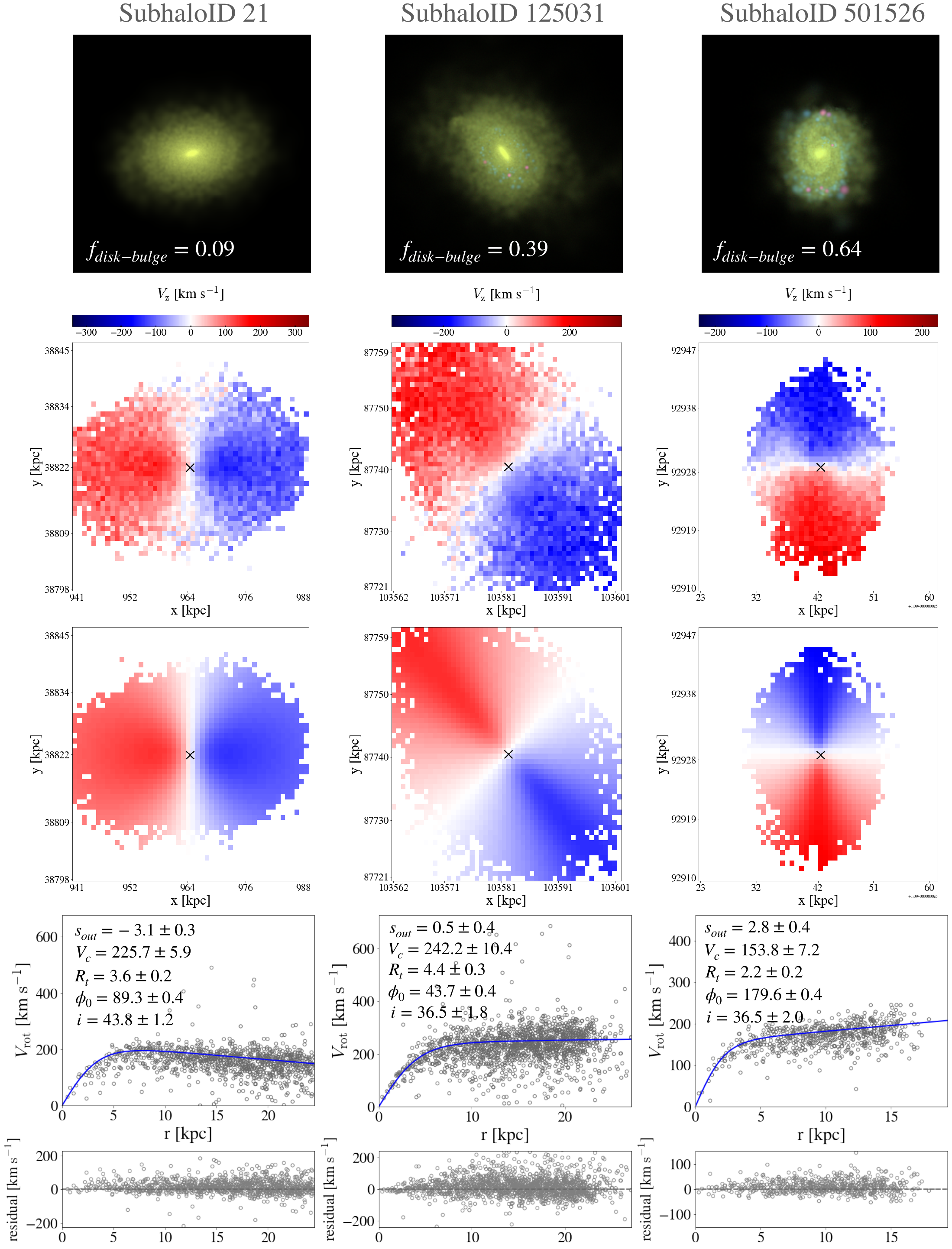}
\caption{Examples of galaxies in this study. The first row: color SKIRT synthetic images. The SubhaloID on the top is the subhalo index that is assigned by the SUBFIND algorithm. The $f_\mathrm{disk-bulge}$ on the lower left corner is the dynamical disk fraction without the contribution of the "bulge" to the "disk", which provides insight into the dynamical morphology of the galaxy. The second row: 2D mock line-of-sight velocity maps. The third row: the best-fit 2D model maps for the upper velocity maps. The bottom row: 1D velocity profiles. The circles show rotational velocities at r (distance from the kinematic center) after correcting for the geometric term cos($\phi$ -$\phi_0$) and the kinematic inclination sin $i$ (shown here are data points located within ± 60° from the kinematic major axis). The blue solid lines are the best-fit rotation curve models of Equation (\ref{eq:Vr}). In each panel of the bottom row, we provide the fitted model parameters of Equations (\ref{eq:Vr}) and (\ref{eq:Vobs}). Bottom panels show residual velocities ($V_\mathrm{model} - V_\mathrm{obs}$). \label{fig:RC_early}}
\end{figure*}

\begin{figure}[t!]
\centering
\includegraphics[width=80mm,scale=1]{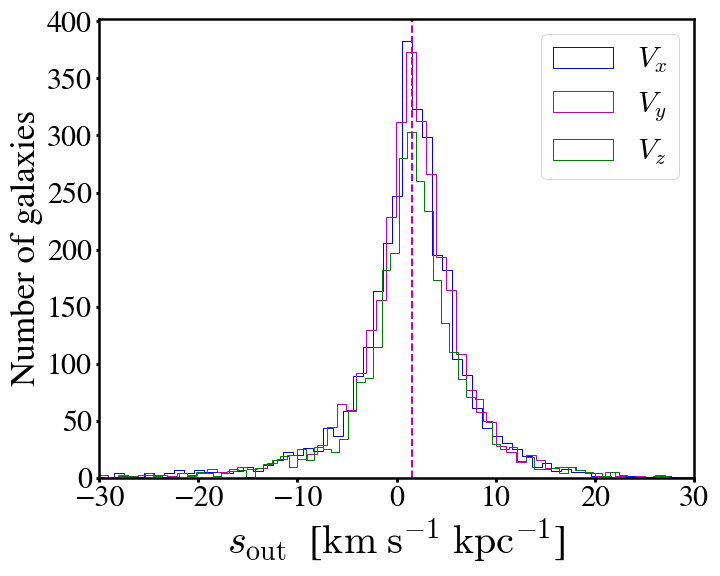}
\caption{Distributions of $s_\mathrm{out}$ of final sample galaxies in different lines of sight. The blue, magenta, and green histograms show the distribution of final samples on X-, Y-, and Z-axis line-of-sight velocity, respectively. The magenta dashed vertical line shows the median value of the $s_\mathrm{out}$ distribution of $V_\mathrm{y}$ line of sight ($s_\mathrm{out}$ =1.6).     \label{fig:southist}}
\end{figure}

\begin{figure*}[hbt]
\centering
\includegraphics[width=180mm,scale=1]{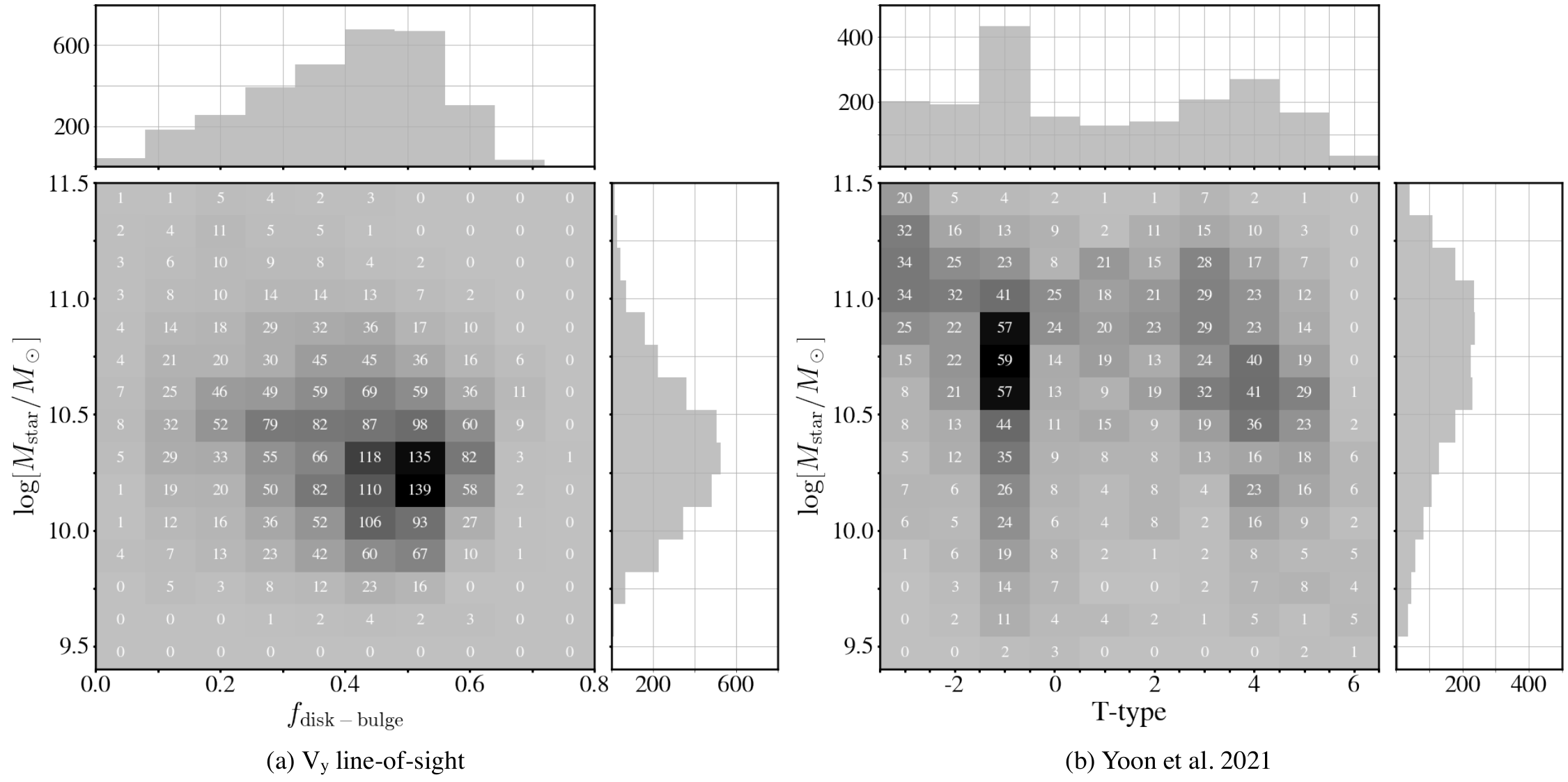}
\caption{The number of galaxies utilized to construct the Figure \ref{fig:mainresult} in the Morphology versus log $M_\mathrm{star}$ plane. The galaxy counts are represented by grey level and numbers in each bin. The left figure depicts the result of simulation on $V_\mathrm{y}$ line-of-sight velocity and the right figure depicts the result of observation in \citet{yoon2021rotation}. We use a grid in which block sizes in the $f_\mathrm{disk-bulge}$ and $\log{M_\mathrm{star}}$ axes are 0.08 and 0.14, respectively. The top and right panels of each figure show 1D histograms of $f_\mathrm{disk-bulge}$ and $\log{M_\mathrm{star}}$, respectively. \label{fig:2Dhist}}
\end{figure*}

\begin{figure*}[hbt!]
\centering
\includegraphics[width=160mm,scale=1]{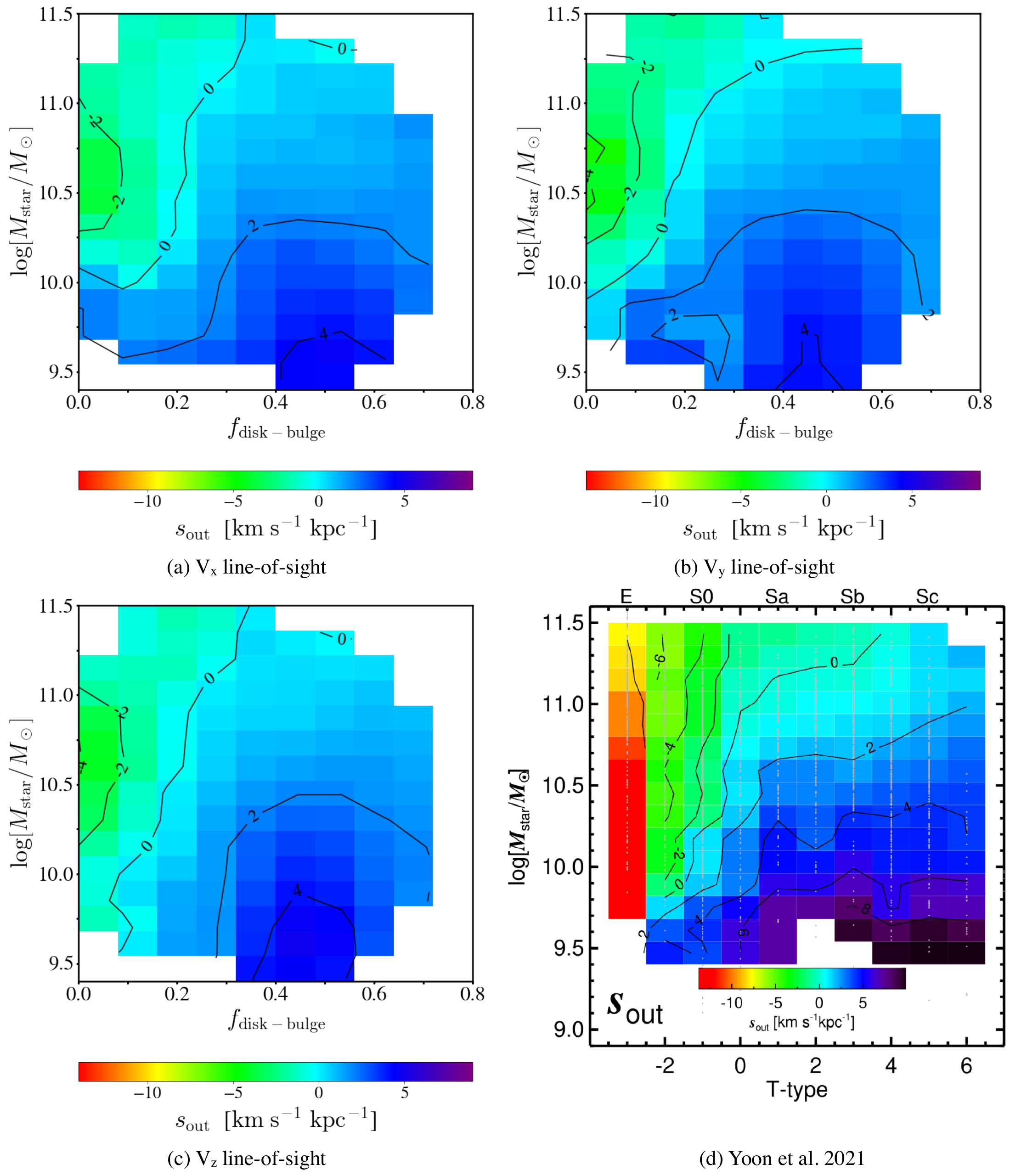}
\caption{The outer slopes of rotation curves ($s_\mathrm{out}$) in the Morphology versus log $M_\mathrm{star}$ plane. The outer slopes ($s_\mathrm{out}$) are represented by colors (see the color bars for the color-coded $s_\mathrm{out}$) and contours (values in the middle of contour lines indicate $s_\mathrm{out}$). Upper left, upper right, and lower left figures depict the results on $V_\mathrm{x}$, $V_\mathrm{y}$, and $V_\mathrm{z}$ line-of-sight velocity, respectively. To construct color maps and contours for $s_\mathrm{out}$, we use a grid in which block sizes in the $f_\mathrm{disk-bulge}$ and $\log{M_\mathrm{star}}$ axes are 0.08 and 0.14, respectively. At each point in the grid, we calculated the median slopes for galaxies within a rectangular bin whose sizes in the T-type and $\log{M_\mathrm{star}}$ sides are 0.16 and 0.7, respectively. We display only the 2D bins that contain a minimum of 20 galaxies. For more details about the right panel, please refer to the caption of Figure 5 in \citet{yoon2021rotation}. \label{fig:mainresult}}
\end{figure*}

\begin{figure*}[hbt!]
\centering
\includegraphics[width=160mm,scale=1]{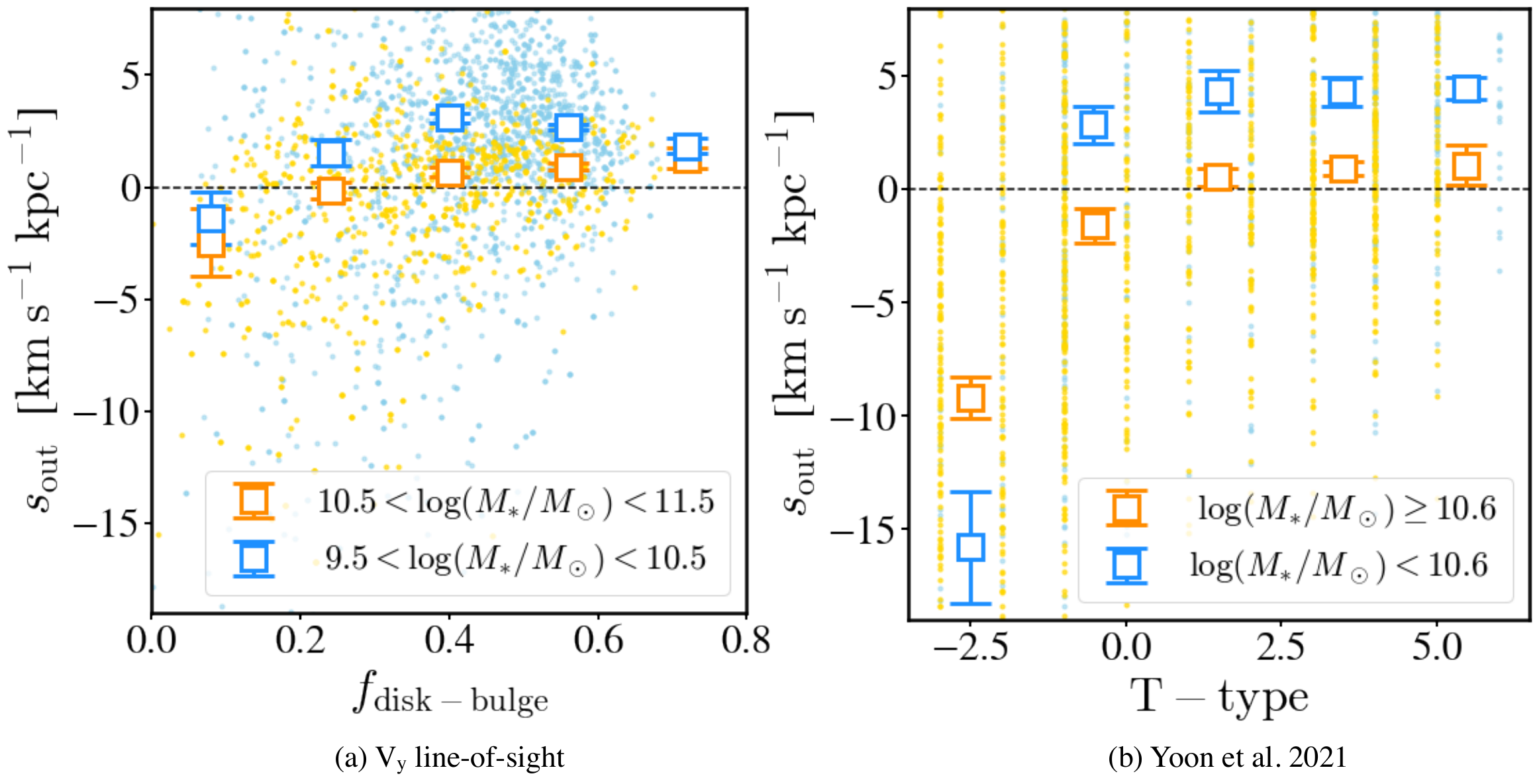}

\caption{The outer slopes of rotation curves ($s_\mathrm{out}$) in the morphology. The final results of $V_\mathrm{y}$ line of sight are used. The more massive samples ($10.5 < \log{[M_\mathrm{star}/M_\odot]} < 11.5$) and less massive samples ($9.5 < \log{[M_\mathrm{star}/M_\odot]} < 10.5$) are presented in orange and blue color, respectively. The square denotes the median value within a horizontal bin whose size is 0.16 in the disk fraction. We display only the data points that contain a minimum of 5 galaxies. The error bar is the standard deviation of the median values from 1000 bootstrap resamplings. Small dots are individual galaxies. For more details about the right panel, please refer to the caption of Figure 6 in \citet{yoon2021rotation}.}   \label{fig:sout_morph}
\end{figure*}

\begin{figure*}[hbt!]
\centering
\includegraphics[width=160mm,scale=1]{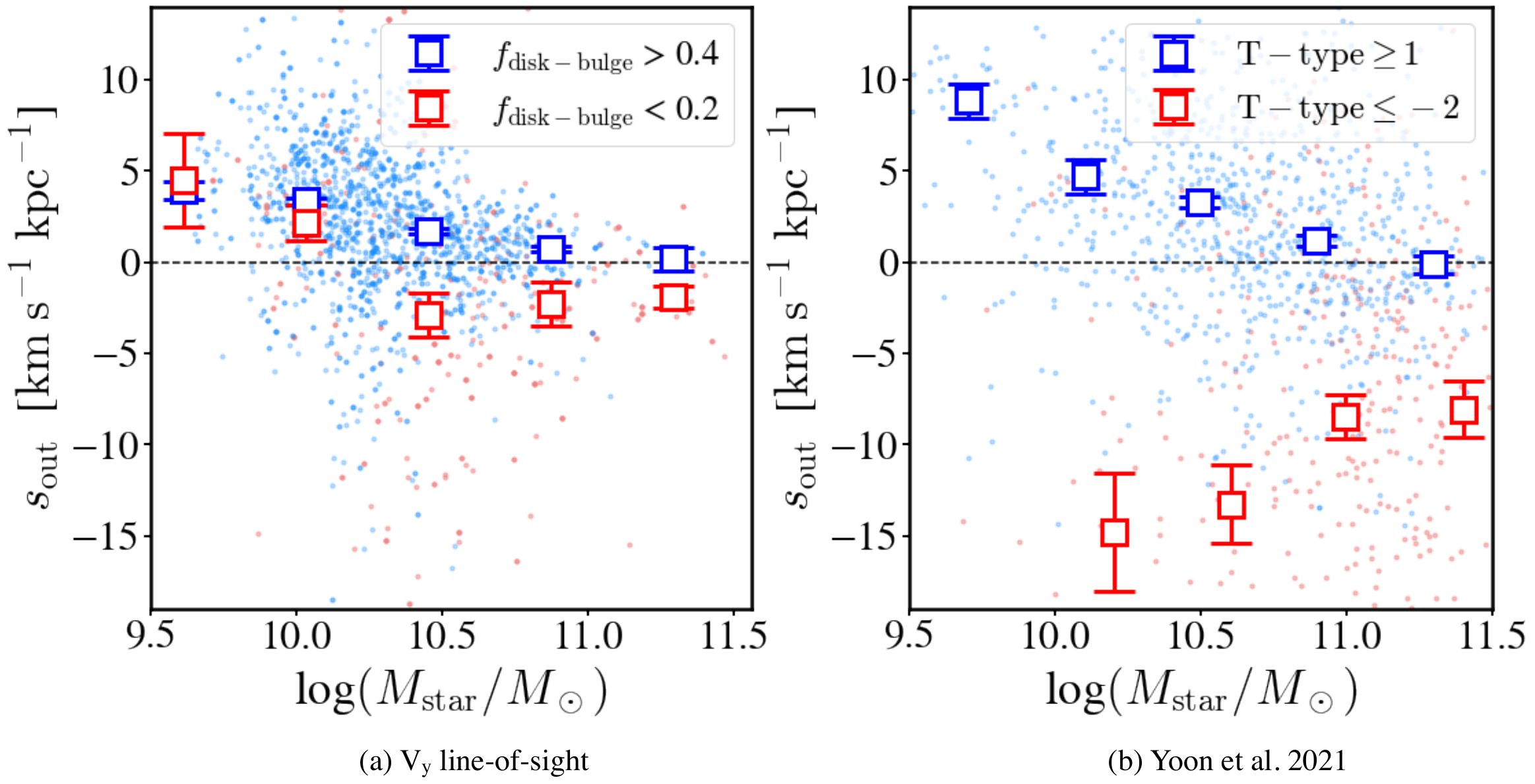}

\caption{The outer slopes of rotation curves ($s_\mathrm{out}$) in the stellar mass of the galaxies. The final results of $V_\mathrm{y}$ line of sight are used. The early-type samples ($f_\mathrm{disk-bulge}<0.2$) and late-type samples ($f_\mathrm{disk-bulge}>0.4$) are presented in red and blue color, respectively. The square denotes the median value within a bin whose size is 0.42 in the $\log{(M_\mathrm{star}/M_\odot)}$. We display only the data points that contain a minimum of 5 galaxies. The error bar is the standard deviation of the median values from 1000 bootstrap resamplings. Small dots are individual galaxies. For the details of (b) Yoon et al. 2021, refer to the caption of Figure 6 in \citet{yoon2021rotation}.}  \label{fig:sout_mass}
\end{figure*}

\section{Results} \label{sec:results}
 
In this section, we present the key findings in our analysis of the dependence of rotation curve shape on the stellar mass and on the morphology. Figure \ref{fig:RC_early} shows our analysis for three representative galaxies in the final samples. The SubhaloID at the top refers to the subhalo index assigned by the SUBFIND algorithm in IllustrisTNG. The top panels show the color image of each galaxy created with the SKIRT radiative transfer code (\citealp{Rodriguez_Gomez_2018}). To obtain the color image, we combine g, r, i broadband filter images similar to SDSS using the $make\_lupton\_rgb$ function in astropy. 

The bottom left corner of the image lists $f_\mathrm{disk-bulge}$ that is the dynamical disk fraction without the effect of the bulge representing the dynamical morphology of the galaxy. \citet{Genel15} calculate stellar circularities of stellar particles in TNG subhaloes using the information of angular momentum and define $f_\mathrm{disk-bulge}$ as the fractional mass of stars with circularity larger than 0.7 minus the fraction of stars with circularity smaller than -0.7. Here, the reversed moving particles with circularity smaller than -0.7 were excluded to remove the contamination of bulge stars (that tend to show symmetric distribution of the circularity parameter) in the selection of disk stars with the circularity larger than 0.7 (see the bottom inset of Fig. 2 in \citealp{Abadi03}). Then, $f_\mathrm{disk-bulge}$ becomes the fraction of stellar particles with a significant positive rotational support without the influence of the bulge. We use the $f_\mathrm{disk-bulge}$ as a proxy for galaxy morphology instead of morphological type purely based on images (e.g. the T-type system used in \citealp{yoon2021rotation}). Comparisons across different morphologies are discussed further in Section \ref{subsec:morph}. 


The second row in Figure \ref{fig:RC_early} shows the two-dimensional velocity map of each galaxy constructed with the method described in Section \ref{subsec:Vmap}. The third row represents the best model fit of the corresponding velocity map in the second row derived with Equations (\ref{eq:Vr}) and (\ref{eq:Vobs}). Because we have only the synthetic SKIRT images on x-y plane, the velocity maps in Figure \ref{fig:RC_early} show the $V_\mathrm{z}$ line-of-sight velocity on x-y plane. The bottom row describes the fitted one-dimensional rotation curves of galaxies with ($V_\mathrm{model}-V_\mathrm{obs}$) residuals. The data points represent the rotational velocities at a given distance (r) from the kinematic center of the galaxy. These velocities have been corrected for the geometric term $\cos{(\phi - \phi_0)}$, which accounts for the angle between the line connecting each data point to the galaxy center and the kinematic major axis of the galaxy. Additionally, the velocities have been adjusted for the kinematic inclination of the galaxy with $\sin{i}$. The data points shown here have a position angle within $\pm 60^\circ$ from the kinematic major axis (i.e. close to major axis). There are the fitted model parameters of Equations (\ref{eq:Vr}) and (\ref{eq:Vobs}) ($V_c$ in km s$^{-1}$, $R_\mathrm{t}$ in kpc, $s_\mathrm{out}$ in km s$^{-1}$ kpc$^{-1}$, and $i$ and $\phi_0$ in degrees) on the top left corner of the bottom panels. 

The fitted results in both 2D and 1D plots appear satisfactory overall, exhibiting a good agreement with the mock observed data. However, there might be some minor variations with scatters, which could be attributed to measurement uncertainties, noise, or intrinsic complexities in the rotation curves of individual galaxies. Nevertheless, the general trends and shapes of the rotation curves are well captured by the fitting procedure. The functional form of Equation (\ref{eq:Vr}) represents and quantifies well the rotation curves of galaxies. This means that we can analyze the galaxy rotation curves with Equations (\ref{eq:Vr}) and (\ref{eq:Vobs}) as long as galaxies have a sufficient amount of rotation even for early-type galaxies (\citealp{Capellari16, Graham18}). 

Figure \ref{fig:southist} displays the distributions of $s_\mathrm{out}$ of final samples in different lines of sight. The histograms in blue, magenta, and green depict the distribution of final samples along the X-axis, Y-axis, and Z-axis line-of-sight velocities, respectively. All the histograms shows a peak around $s_\mathrm{out}$=1.6.

Figure \ref{fig:2Dhist} shows the distribution of galaxies in the plane of morphology and stellar mass. The galaxy counts are displayed with both grey level and number in each bin. The sample in the simulation (left panel) exhibits a peak around $f_\mathrm{disk-bulge}$=0.5 and $\log(M_\mathrm{star}/M_\odot)$=10.3. The distribution of the observation sample in \cite{yoon2021rotation} has a bimodal feature for T-type morphology, with a peak occurring at approximately $10^{11.0}M_\odot$ of the stellar mass.

Figure \ref{fig:mainresult} shows the dependency of the outer slope of the rotation curve ($s_\mathrm{out}$) on the stellar mass and on the morphology of a galaxy. The three results with different lines of sight have generally similar trends as expected. The outer slope ($s_\mathrm{out}$) decreases as galaxies are more massive, and increases as galaxies are more disky. The early-type galaxies with a stellar mass around $10^{10.5}M_\odot$ have the smallest $s_\mathrm{out}$ value for all the lines of sight. Because the sample for each line of sight is slightly different, the results from three lines of sight are not exactly the same. The comparison with observational result (i.e. bottom right panel) is shown in Section \ref{subsec:observ}.  

Figure \ref{fig:sout_morph} provides a clear representation of the changes in $s_\mathrm{out}$ with galaxy morphology. The samples are divided into two: more massive ($10.5 < \log{[M_\mathrm{star}/M_\odot]} <11.5$) and less massive ($9.5 < \log{[M_\mathrm{star}/M_\odot]} < 10.5$) galaxies. Both samples tend to have larger $s_\mathrm{out}$ as $f_\mathrm{disk-bulge}$ is larger except for the rotation-dominated (i.e. $f_\mathrm{disk-bulge} >0.5$), less-massive galaxies. When $f_\mathrm{disk-bulge} >0.5$, the median $s_\mathrm{out}$ tends to decrease slightly or to remain flat for the less massive galaxies. The $s_\mathrm{out}$ for less massive galaxies ranges from -1 to 3 km s$^{-1}$ kpc$^{-1}$, and then settles around 1.8 km s$^{-1}$ kpc$^{-1}$. The more massive galaxies have $s_\mathrm{out}$ ranging from -2 to 1 km s$^{-1}$ kpc$^{-1}$. The $s_\mathrm{out}$ of more massive galaxies is always smaller than that of less massive galaxies for any $f_\mathrm{disk-bulge}$. The comparison with observational result (i.e. right panel) is shown in Section \ref{subsec:observ}.

Figure \ref{fig:sout_mass} displays how $s_\mathrm{out}$ changes with the stellar mass of galaxies. The samples are categorized into two groups based on their morphology: early-type galaxies with $f_\mathrm{disk-bulge} < 0.2$ and late-type galaxies with $f_\mathrm{disk-bulge}>0.4$ (roughly corresponding to  T-type $\leq$ -2 and T-type $>$ 2, respectively). For late-type galaxies, the outer slope tends to decrease from around 4 to 0 ($\Delta s_\mathrm{out} = $ -4 km s$^{-1}$ kpc$^{-1}$) with increasing stellar mass. The early-type galaxies show the minimum value of $s_\mathrm{out}$ ($s_\mathrm{out} \approx $ -3 km s$^{-1}$ kpc$^{-1}$) at stellar mass around $10^{10.5}M_\odot$. The early-type galaxies, on average, show smaller values of $s_\mathrm{out}$ than late-type galaxies. 

\begin{figure*}[hbt!]
\centering
\includegraphics[width=150mm,scale=1.2]{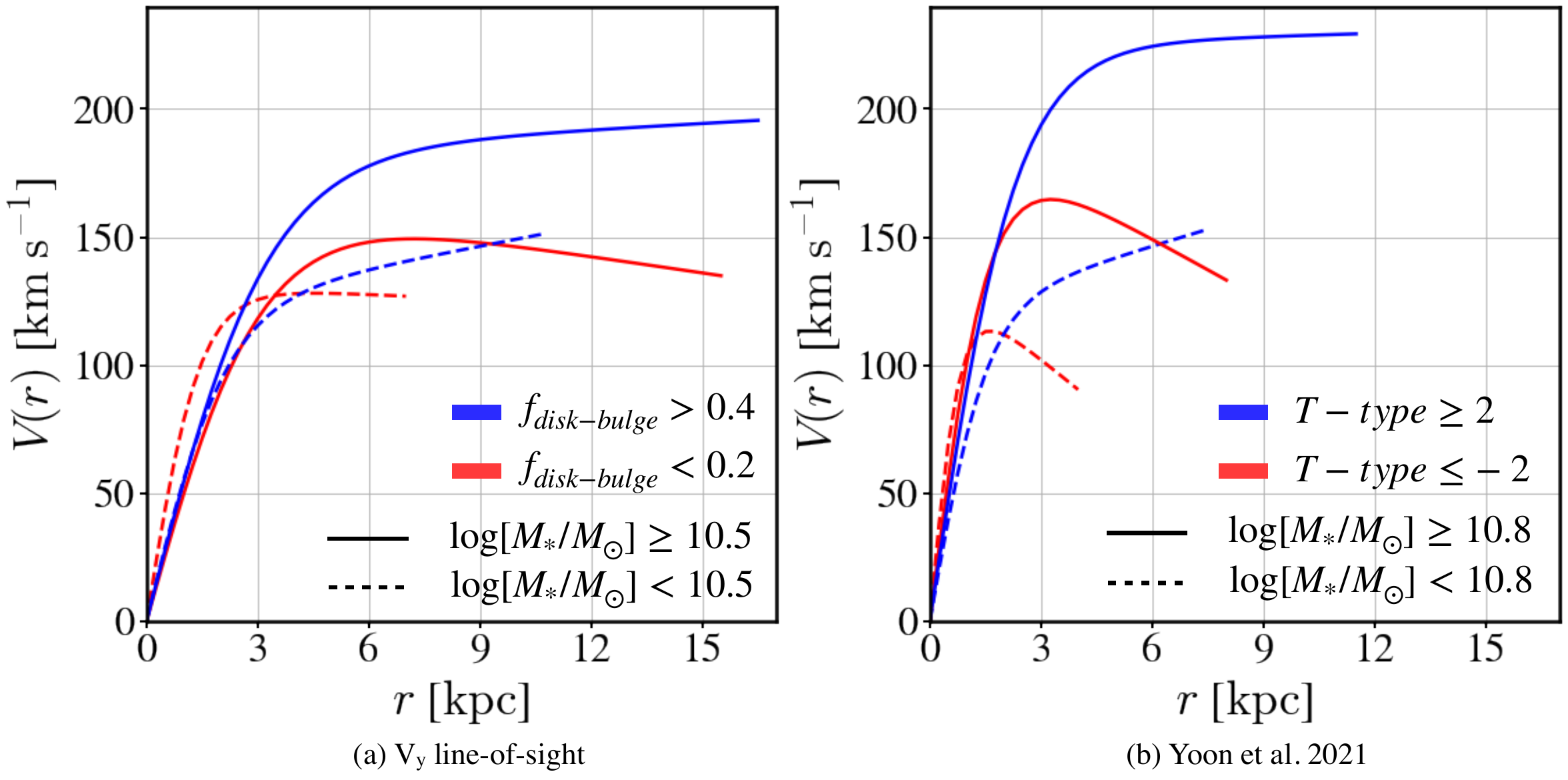}

\caption{Typical rotation curves of galaxies with various morphology and stellar mass. The final results of $V_\mathrm{y}$ line of sight are used. The rotation curves are generated by calculating the median values of rotation curve parameters (as defined in equation \ref{eq:Vr}) for each subsample. The resulting rotation curve profiles are displayed up to 5 times the turnover radius ($5R_\mathrm{t}$)\label{fig:typicalRC}}
\end{figure*}

\begin{table*}[t!]
  \centering
  \caption{The values of the RC parameters for typical rotation curves: $s_\mathrm{out}$ [km s$^{-1}$ kpc$^{-1}$], $V_c$ [km s$^{-1}$], and $R_\mathrm{t}$ [kpc] in Equation (\ref{eq:Vr}). Figure \ref{fig:typicalRC} shows the high  and low mass ($M_*$) samples with solid and dashed lines, respectively. The late-type and early-type samples are represented by blue and red colors, respectively. The error is the standard deviation of the median values from 1000 bootstrap resamplings. \label{table:typicalRC}}
  \begin{tabular}{l c c c c c c}
    \toprule
    Sample & \multicolumn{3}{c}{(a) This Work} & \multicolumn{3}{c}{(b) Yoon et al. 2021} \\
    \midrule
           & $s_\mathrm{out}$ & $V_c$ & $R_\mathrm{t}$ & $s_\mathrm{out}$ & $V_c$ & $R_\mathrm{t}$ \\
    \cmidrule(lr){2-4}\cmidrule(lr){5-7}
    High $M_*$, Late-type  & 0.8$\pm$0.1  & 182.5$\pm$2.6 & 3.3$\pm$0.1  & 0.5$\pm$0.3  & 224.1$\pm$5.8  & 2.4$\pm$0.1 \\
    High $M_*$, Early-type  & -2.1$\pm$0.5 & 168.2$\pm$7.0 & 3.2$\pm$0.3  & -8.1$\pm$1.0 & 196.1$\pm$10.8 & 1.7$\pm$0.1 \\
    Low $M_*$, Late-type   & 2.8$\pm$0.1  & 121.4$\pm$1.3 & 2.2$\pm$0.0  & 3.8$\pm$0.3  & 121.0$\pm$3.5  & 1.6$\pm$0.0 \\
    Low $M_*$, Early-type  & -0.7$\pm$0.7 & 131.6$\pm$5.7 & 1.4$\pm$0.1  & -14.3$\pm$1.9& 139.8$\pm$13.1 & 0.9$\pm$0.0 \\
    \bottomrule
  \end{tabular}
\end{table*}

Figure \ref{fig:typicalRC} shows the rotation curves of galaxies with different morphological types ($f_\mathrm{disk-bulge}$) and stellar masses ($M_\mathrm{star}$) generated by taking the median values of the rotation curve parameters of Equation (\ref{eq:Vr}) for each subsample. Late-type galaxies with a larger value of $f_\mathrm{disk-bulge}$ are represented by blue color, while early-type galaxies with a smaller value of $f_\mathrm{disk-bulge}$ are represented by red color. More massive galaxies are denoted by solid lines, and less massive galaxies are denoted by dashed lines. Late-type galaxies show flat or slightly increasing rotation curves at the outer part, while early-type galaxies show decreasing shape of rotation curves. In general, late-type galaxies and more massive galaxies tend to have larger values of both the velocity ($V_c$) and the turnover radius ($R_\mathrm{t}$) than early-type galaxies and less massive galaxies. This means that late-type galaxies and more massive galaxies exhibit stronger rotational motion and larger physical sizes in terms of their rotation curves.


\section{Discussion} \label{sec:discussion}

\subsection{Comparison with Observations} \label{subsec:observ}

Here, we discuss our results from simulations in comparison with observations. In Figure \ref{fig:mainresult}, where we examine the dependence of outer slope of galaxy rotation curves on stellar mass and on morphology, it should be noted that the number of samples is not large enough in the highest (i.e. $M_\mathrm{star}>10^{11.0}M_\odot$) and lowest ($M_\mathrm{star}<10^{9.75} M_\odot$) mass bins (See Figure \ref{fig:2Dhist}). For the highest mass bins, the limited volume of the simulation box would be the main reason why the number of galaxies is small. The observational samples of MaNGA are in the redshift range of z $<$ 0.15 (\citealp{Wake17}). The comoving distance of z = 0.15 is about 592 Mpc. However, the volume of the simulation box is (106.5 Mpc)$^3$ which is much smaller than that of the observation. The limited volume of the simulation would have prevented the formation of galaxies with large mass created by merging process. In the case of the lowest mass bins, the number of galaxies is small because the chance to fail the rotation curve fit is high for low-mass galaxies where the mass resolution effect in simulations is significant. The stellar particles in TNG 100 have masses of approximately $10^6M_\odot$, which is much larger than the actual mass of a single star. To get rid of noisy pixels, we use only the pixels with the stellar particles larger than nine in the 2D velocity map for RC fitting. In this process, the outer regions of the velocity map are eliminated for the galaxies with low masses, resulting in only the central part of the galaxy remaining for analysis. Therefore, many low-mass galaxies without reliable fitting results are filtered out in Section \ref{subsec:RCfitting} because there are no data points in the outer part for the fit.

In Figures \ref{fig:mainresult}-\ref{fig:typicalRC}, the overall trend of $s_\mathrm{out}$ from this work using simulations is similar to the result of \citet{yoon2021rotation}. The outer slope increases as galaxies are more disky, and decreases as galaxies are more massive. The consistent results between simulations and observations suggest that we can use the simulation data for further analysis to understand the physical origin of diverse rotation curves.  

Although the overall trend of $s_\mathrm{out}$ follows the result of the observation, there are some differences between observations and simulations. Figures \ref{fig:mainresult} and \ref{fig:sout_mass} show that there is a notable discrepancy for the value of $s_\mathrm{out}$ between observations and simulations for early-type galaxies, with the former exhibiting significantly lower values. Moreover, the $s_\mathrm{out}$-mass dependence for early-type galaxies differs between observations and simulations. In observations, $s_\mathrm{out}$ of early-type galaxies becomes larger with stellar mass. However, $s_\mathrm{out}$ in simulations first decreases with stellar mass, and increases with stellar mass at $M_\mathrm{star}>10^{10.5}M_\odot$. These differences could be understood as follows.

First, the sample difference between observations and simulations could be a cause. We find that the galaxies with merging features exhibit significantly lower $s_\mathrm{out}$ than those without merging features. Also, the galaxies with merging features would have lower $f_\mathrm{disk-bulge}$ because they exhibit a certain degree of non-circular motion. Therefore, we think that a significant exclusion of such galaxies in simulations would be the main reason for the difference in $s_\mathrm{out}$ for early-type galaxies between observations and simulations. Furthermore, the lack of high-mass early-type galaxies in the simulation as in Figure \ref{fig:2Dhist} could have affected the trend of $s_\mathrm{out}$.

Second, the discrepancy could be induced by the use of different proxy for galaxy morphology: T-type and $f_\mathrm{disk-bulge}$. The visually classified galaxy morphology (T-type) and dynamically derived morphology ($f_\mathrm{disk-bulge}$) do not always agree each other. For example, there are also fast rotators in ellipticals (\citealp{Capellari16, Lee18}). Elliptical galaxies that exhibit significant rotation may not be classified as early-type galaxies based on the dynamical morphology parameter ($f_\mathrm{disk-bulge}$). This difference could potentially contaminate the galaxy sample based only on optical images.

Third, the discrepancy could result from the limits of TNG simulations. As we will discuss in next section, the dark matter fraction in a galaxy is correlated with the outer slope of the rotation curve. According to the findings of \citet{SEAGLE22}, the dark matter fraction of TNG galaxies within the half effective radius is not consistent with the observational measurements. They propose that the primary factor contributing to this discrepancy is likely the stellar feedback model employed in the TNG simulations. Stellar feedback plays a crucial role in shaping the distribution of baryonic matter including stars and gas inside galaxies. The specific implementation of stellar feedback in the simulation can influence the overall distribution and properties of dark matter, resulting in the change of $s_\mathrm{out}$. 

In both Figures \ref{fig:mainresult} and \ref{fig:sout_morph}, the result from observations shows that $s_\mathrm{out}$ remains relatively constant after an initial increase with T-type from -2.5 to 2. However, our results from simulations show that the $s_\mathrm{out}$ at $f_\mathrm{disk-bulge}>0.7$ appears slightly smaller than that at $f_\mathrm{disk-bulge} \approx 0.6$ when the stellar mass is smaller than $10^{10.5}M_\odot$. This may also result from a different definition of the morphology. Galaxies with T-type values of 5 or 6 are typically classified as Sc or Scd galaxies, which are characterized by an ill-defined or loose spiral structure. These galaxies may have comparatively lower  $f_\mathrm{disk-bulge}$ because they often exhibit less pronounced spiral arms than earlier spiral types such as Sa or Sb galaxies.

Figures \ref{fig:sout_morph}-\ref{fig:typicalRC} show that the range of $s_\mathrm{out}$ in this work is narrower than that of \citet{yoon2021rotation}. This may come from the deficit in samples of highest and lowest masses. Additionally, the slightly different fitting range may have influenced the final value of $s_\mathrm{out}$. As discussed in Section \ref{subsec:Vmap}, the radial coverage of 2D velocity map in this work is slightly larger than that of the observation. If the true outer rotation curve follows a quadratic function instead of a linear function, the absolute value of the outer slope could be measured to be smaller when flat outer regions are more included in the analysis. 

In Figure \ref{fig:typicalRC} and Table \ref{table:typicalRC}, $V_c$ of more massive galaxies in simulations are smaller than those in observations. This difference would be primarily attributed to the limited volume of the simulation box, which results in the lack of highest-mass galaxies ($M_{*}>10^{11.0}M_\odot$) in the simulation sample. The values of $R_\mathrm{t}$ are about twice larger than those of observations. The physical cause of the difference is unknown. 

In conclusion, the results from simulations show various trends of $s_\mathrm{out}$ that are roughly consistent with observations; however, there are some details that are different between the two, which requires some caution.


\subsection{Dark Matter fraction and Gas fraction} \label{subsec:DMF}

\begin{figure*}[hbt!]
\centering
\includegraphics[width=160mm,scale=1]{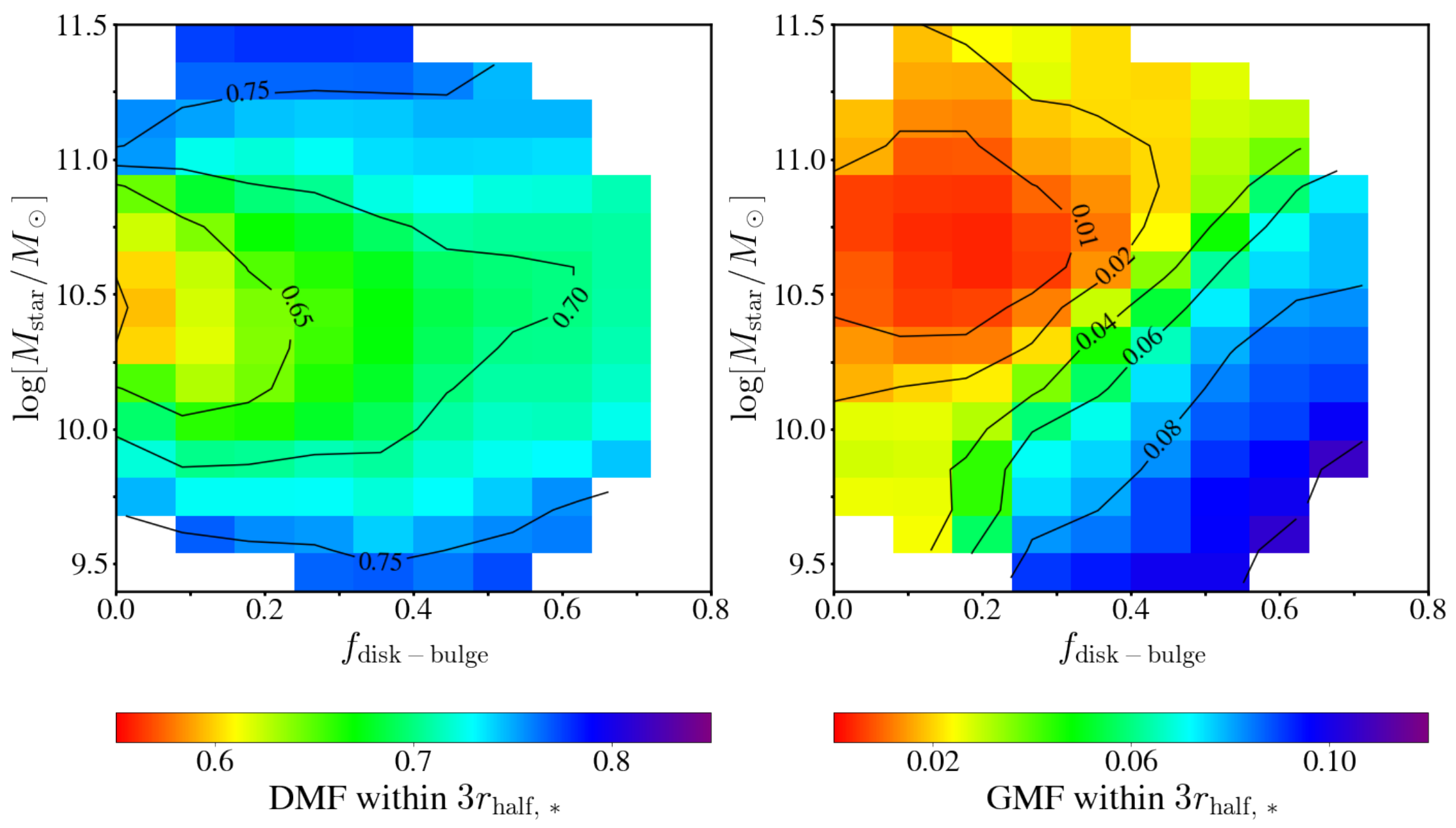}
\caption{Dark matter mass fraction (Left panel) and Gas mass fraction (Right panel) in the Morphology versus log $M_\mathrm{star}$ plane. The final results of $V_\mathrm{y}$ line of sight are used. Dark matter fraction and gas fraction represented by colors (see the color bars) and contours (values in the middle of contour lines). Other descriptions are identical to those in Figure \ref{fig:mainresult}. \label{fig:DMFGasF}}
\end{figure*}

\begin{figure*}[t!]
\centering
\includegraphics[width=160mm,scale=1]{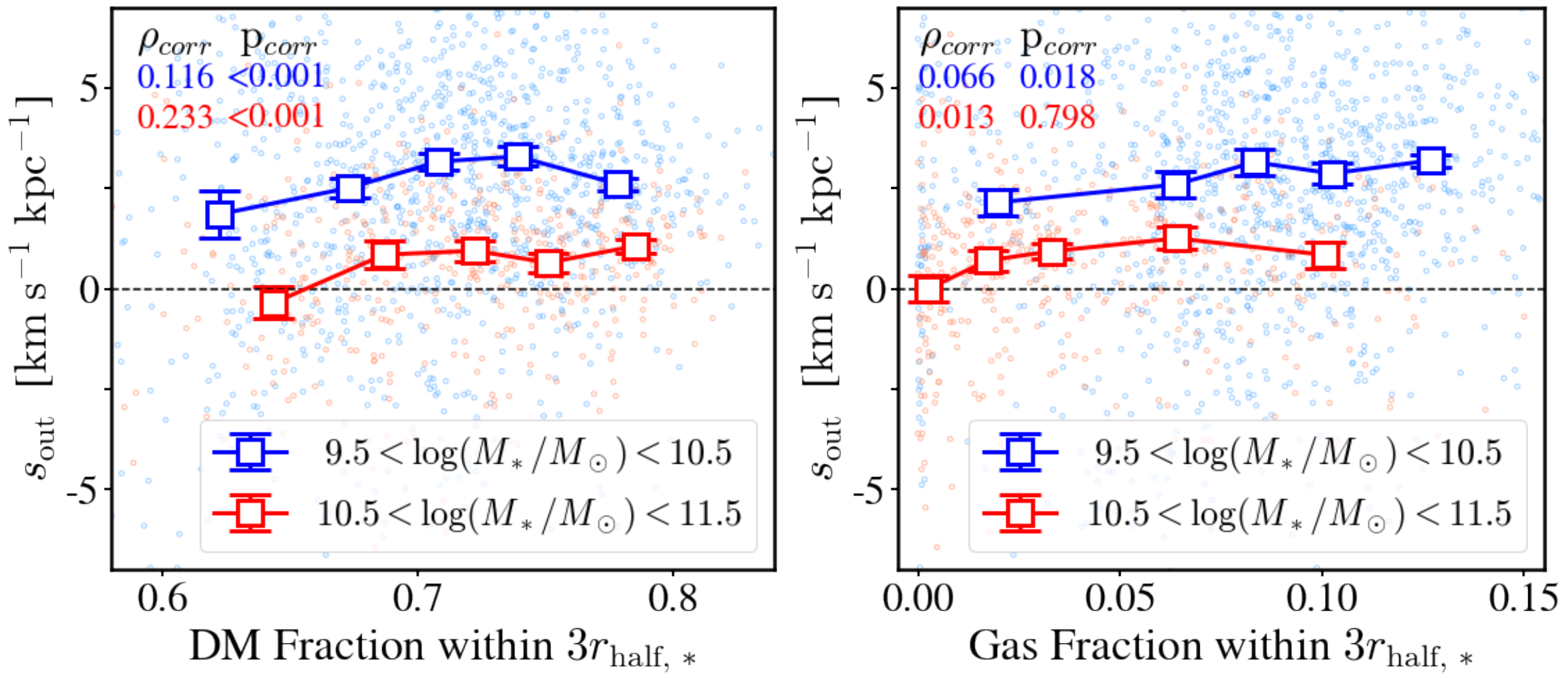}
\caption{The outer slopes of rotation curves ($s_\mathrm{out}$) as a function of the dark matter fraction (Left panel) and the gas fraction (Right panel) of the galaxies. Only the late-type galaxies ($f_\mathrm{disk-bulge}>0.4$) with strong rotation in $V_\mathrm{y}$ line of sight are utilized. The less massive and more massive galaxies are presented in red and blue color, respectively. Small open circles are individual galaxies. The square denotes the median value within a bin equalized by the number of galaxies entering each bin. The number of galaxies in each bin is 256 or 257 for less massive galaxies, and 82 or 83 for more massive galaxies. The error bar is the standard deviation of the median values from 1000 bootstrap resamplings. The Pearson correlation coefficient (PCC) values, computed using data from individual galaxies, are displayed at the top of the figures based on the mass of the samples.     \label{fig:DMF_GasF_sout}}
\end{figure*}

To understand the underlying physical factors contributing to the changes in the shape of the rotation curves, we examine the dark matter mass fraction (DMF) and the gas mass fraction (GMF). The mass fractions of dark matter and gas are derived as 

\begin{equation}
    DMF = \frac{M_\mathrm{DM, 3r_\mathrm{half,*}}}{M_\mathrm{tot, 3r_\mathrm{half,*}}}, GMF = \frac{M_\mathrm{Gas, 3r_\mathrm{half,*}}}{M_\mathrm{tot, 3r_\mathrm{half,*}}},
\end{equation}
where $M_\mathrm{DM (Gas), 3r_\mathrm{half,*}}$ is the sum of the dark matter (gas) particles of the galaxy within three times the stellar half-mass radius, and $M_\mathrm{tot, 3r_\mathrm{half,*}}$ is the sum of all the particles of the galaxy within the same radius.

Figure \ref{fig:DMFGasF} shows the mass fractions of dark matter and gas as functions of morphology ($f_\mathrm{disk-bulge}$) and stellar mass ($\log{[M_\mathrm{star}/M_\odot]}$) for the $V_\mathrm{y}$ line of sight. The comparison between Figures \ref{fig:mainresult} and \ref{fig:DMFGasF} suggests that there is a notable similarity between the dark matter fraction and $s_\mathrm{out}$; the smallest $s_\mathrm{out}$ appears for early-type galaxies with a stellar mass around $10^{10.5} M_\odot$. This result is consistent with previous finding for the correlation between the dark matter fraction and stellar mass in TNG subhaloes (\citealp{Lovell18}). On the other hand, the gas fraction in the right panel of Figure \ref{fig:DMFGasF} shows a negative peak located at a higher stellar mass and larger disk fraction than the case of $s_\mathrm{out}$.

Figure \ref{fig:DMF_GasF_sout} shows the outer slopes of rotation curves ($s_\mathrm{out}$) depending on dark matter fraction and gas fraction. The square shows the median values of the slope of the outer region of the rotation curve ($s_\mathrm{out}$) as a function of the dark matter (gas) fraction. We equalize the number of galaxies within each bin to obtain median $s_\mathrm{out}$ with similar significance. The tiny open circles in the background are individual galaxies. The subsamples, distinguished by different colors, are categorized according to the stellar mass of the galaxies: less massive galaxies are represented in blue, while more massive galaxies are depicted in red. 

In Figure \ref{fig:DMF_GasF_sout}, we use only the galaxies whose disk fraction ($f_\mathrm{disk-bulge}$) is greater than 0.4 to directly examine the correlation between the mass distribution and the shape of the rotation curve only for systems exhibiting strong rotation. This preference arises because the velocity dispersion can affect the shape of the rotation curve (\citealp{Burkert_2010, Lang_2017, genzel2020rotation, yoon2021rotation}). Hence, we select only the galaxies with relatively strong rotation to study which mass component of a galaxy significantly influences the outer shape of the rotation curve.

Comparing the trends of the median values, 
there is a hint that the dark matter fraction has a positive correlation with $s_\mathrm{out}$. However, it is difficult to find a notable correlation between gas fraction and $s_\mathrm{out}$. To quantify the correlations, we derive the Pearson correlation coefficient that measures linear correlation between two sets of data and show them in each panel of Figure 11 (i.e. $\rho_{corr}$). We also derive and show the p-value that is the probability of the Pearson correlation coefficient occurring by chance (i.e. p$_{corr}$). The Pearson correlation coefficients between dark matter fraction and $s_\mathrm{out}$ are 0.12 and 0.23 for less massive and more massive galaxies, respectively. The Pearson correlation coefficients between gas fraction and $s_\mathrm{out}$ are 0.07 and 0.01 for the less massive and more massive galaxies, respectively. These results suggest a relatively stronger correlation between dark matter fraction and $s_\mathrm{out}$. The p-values also indicate that the chance correlation of the gas fraction with $s_\mathrm{out}$ is much higher than for the dark matter fraction, especially for the higher mass samples. This can hint that the dark matter fraction plays an important role in determining the outer slopes of rotation curves ($s_\mathrm{out}$) more than the gas fraction. Although this finding is not surprising (because dark matter has been important to explain the flat rotation curve), it is important that we could quantify the contribution and provide a guide for the direction for further study to understand its physical origin. 



\subsection{The Impact of Morphological Classification on the Result} \label{subsec:morph}
\begin{figure}[b!]
\centering
\includegraphics[width=85mm,scale=1]{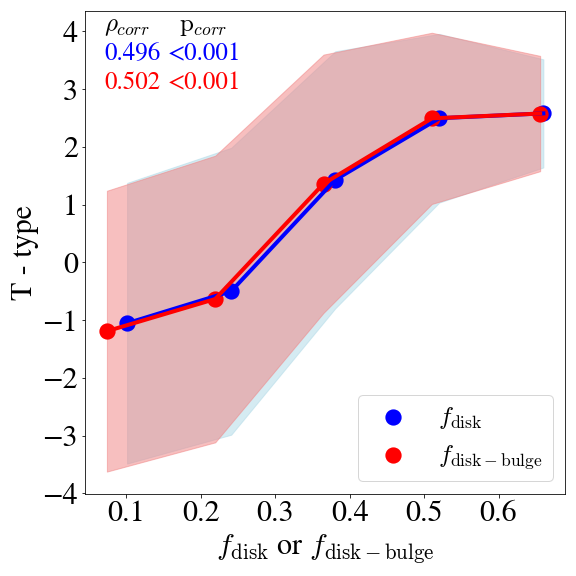}
\caption{The T-type as a function of $f_\mathrm{disk}$ (blue) or $f_\mathrm{disk-bulge}$ (red). The closed circles and shaded regions show the median value and standard deviation of the galaxies, respectively. The Pearson correlation coefficient (PCC) values, computed using data from individual galaxies, are displayed at the top of the figure.}   \label{fig:f-T}
\end{figure}

\begin{figure*}[hbt!]
\centering
\includegraphics[width=160mm,scale=1]{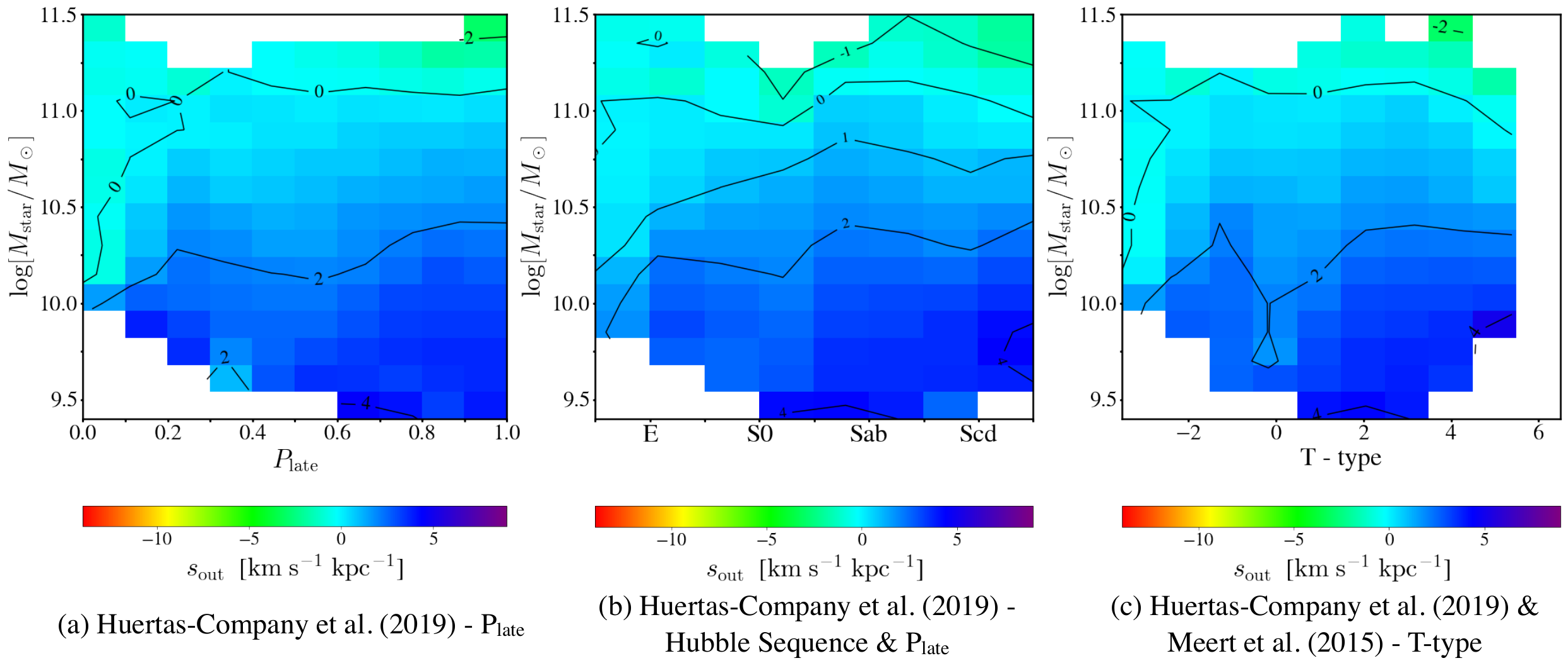}
\caption{The outer slopes of rotation curves ($s_\mathrm{out}$) in the plane of log $M_\mathrm{star}$ and   visually classified galaxy morphology from \citet{HuertasCompany19}. The final results of $V_\mathrm{y}$ line of sight are used. The left panel (a) is based on the morphology of the late-type probability defined by \citet{HuertasCompany19}. The middle panel (b) is the one from the Hubble sequence morphology with the late-type probability. The right panel (c) shows the result from the T-type morphology. The median slope is computed using the horizontal bin sizes of 0.2 for panel (a) and 1.8 for panel (c), respectively. In panel (b), we compute the median slope using the horizontal bin size that includes the bins on both sides of the target bin. Other descriptions are similar to those in Figure \ref{fig:mainresult}.}    \label{fig:diff_morph}
\end{figure*}

We have used the dynamical disk fraction, $f_\mathrm{disk-bulge}$, which is calculated from the stellar circularities of the stellar particles for the morphology of the galaxies (\citealt{Genel15}). This dynamical morphology indicator is not the same as the one based on visual classification in observations. To better understand how our results are affected by the use of different morphology indicators, we compare the morphology indicator from kinematics with those from mock optical images (e.g. visual classification with machine learning from \citealp{HuertasCompany19}), and examine the trends of $s_\mathrm{out}$ with those indicators.


\citet{HuertasCompany19} provide morphological probabilities for galaxies in the IllustrisTNG simulation based on optical images; they use the Convolutional Neural Network (CNN) method trained with the observational images from the Sloan Digital Sky Survey (\citealp{York2000}). To produce SDSS-like images for the simulated galaxies, they use the radiative transfer code, SKIRT (\citealp{Camps15}), and consider the effects of the point spread function and noise. Through three steps of hierarchical binary classification, \citet{HuertasCompany19} offer the late-type probability (i.e. $P_\mathrm{late}=1-P_\mathrm{early}$) and the probabilities of four morphological classes, P(E), P(S0), P(Sab), and P(Scd), for the simulated galaxies.

To obtain the T-type parameter for a simulated galaxy, we adopt the following equation (originally equation (7) of \citealp{Meert15}) that converts the probability from \citet{HuertasCompany19} into the T-type parameter, 
\begin{equation}
\begin{array}{c}
\label{T}
    \mathrm{T} = -\,4.6 \times \, \mathrm{P(E)} -2.4 \times \mathrm{P(S0)} 
\\    +\, 2.5 \times \, \mathrm{P(Sab)} + 6.1 \times \mathrm{P(Scd)}. \end{array}
\label{eq:T}
\end{equation}
This equation is a simple linear model whose coefficients are calibrated to the visually classified galaxies of \citet{Nair&Abraham10} using an unweighted linear regression.

Figure \ref{fig:f-T} shows the relation between the morphological parameters from galactic dynamics ($f_\mathrm{disk}$ and $f_\mathrm{disk-bulge}$) and from T-type classification. Here, $f_\mathrm{disk}$ is the fractional mass of the stars with a circularity greater than 0.7, including the effect of the bulge; this differs from $f_\mathrm{disk-bulge}$ that excludes the effect of the bulge (see Section \ref{sec:results} for details). The plot shows that $f_\mathrm{disk-bulge}$ and $f_\mathrm{disk}$ are almost indistinguishable. Both are moderately correlated to the T-type parameter with the Pearson correlation coefficient of 0.5.


Figure \ref{fig:diff_morph} shows how the outer slope of the rotation curve (i.e. $s_\mathrm{out}$) changes depending on stellar mass and galaxy morphology for the sample of $V_\mathrm{y}$ line of sight. The left panel shows the result based on the late-type probability (i.e. $P_\mathrm{late}$). 
In the middle panel, we first assign the morphology of the galaxy with the highest probability among P(E), P(S0), P(Sab), and P(Scd). Then, we subdivide into two bins per morphology class using late-type probability. The right panel shows the result based on the T-type parameter using Equation (\ref{eq:T}) as in \citet{Nair&Abraham10}.

In the middle panel, \citet{HuertasCompany19} distinguish Ellipticals (E) and Lenticulars (S0) from Spirals (i.e. Sab, Scd) when the late-type probability is smaller than 0.5 ($P_\mathrm{late}<0.5$). Following this philosophy, we subdivide the Ellipticals and Lenticulars in two (one has the late-type probability less than 0.25, and the other one has larger than 0.25), and the Spirals into two (one has the late-type probability less than 0.75, and the other one has larger than 0.75). This is why there are two bins for each class in the middle panel.


The three panels of Figure \ref{fig:diff_morph} show overall similar trends of $s_\mathrm{out}$. We visually inspect the mock images of the galaxies to understand the detailed difference among the three in Figure \ref{fig:diff_morph}. We find that the late-type probability does not match with the Hubble sequence especially for the Scd galaxies. \citealt{HuertasCompany19} classify the late-type spirals and irregulars (T-type $\geq$ 4) as Scd even though they utilize the training set made only of late-type galaxies. It is known that Sd galaxies have diffuse, broken arms. This property could make the late-type probability of Scd galaxies lower than the late-type probability of Sab galaxies.

The overall trend of $s_\mathrm{out}$ in Figure \ref{fig:diff_morph} that is based on galaxy morphology with optical images is similar to the one in Figure \ref{fig:mainresult} that is based on galaxy morphology with kinematics. However, it is difficult to find a clear morphological dependence of $s_\mathrm{out}$ in Figure \ref{fig:diff_morph} compared to Figure \ref{fig:mainresult}; for example, $s_\mathrm{out}$ is not so small for early-type galaxies (i.e. left most column) in Figure \ref{fig:diff_morph} unlike Figure \ref{fig:mainresult}. 

The reason for such a difference could be the resolution limit of simulations, which cannot provide optical images comparable to observed ones. In other words, the spatial resolution of particle-based mock images of galaxies in simulations can make their optical images much clumpier than observations. The limitation is expected to pose challenges in morphological classification. Upon examining galaxies classified as S0 or Scd, those that appear clumpy due to resolution issues present challenges even when seen visually, making their classification difficult. Also, there are quite a few instances where the probabilities of being classified as P(S0) and P(Scd) were comparable.

In conclusion, we try using the optical morphology for the simulated galaxies in TNG to compare equally with the observed galaxies, and find that the results generally agree with that based on galaxy morphology with kinematics (see also \citealp{Foster18}). However, we think that the classification of galaxy morphology with mock optical images is not realistic in this study because of the simulation resolution limit. We therefore think that it is more reasonable to adopt $f_\mathrm{disk-bulge}$ as the main parameter for morphological classification instead of T-type parameter as in observations, which can be directly obtained from simulations with less bias.

\subsection{Circular Velocity Curve} \label{subsec:circ}
\begin{figure}[b!]
\centering
\includegraphics[width=85mm,scale=1]{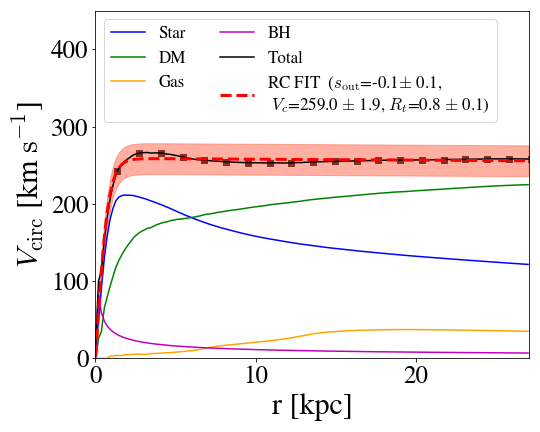}
\caption{An example of the circular velocity curve for the galaxy whose SubhaloID is 125031. It shows the circular velocity curves within three times the stellar half-mass radius. The colored lines indicate the contribution of each component to the circular velocity. The black line with square represents the circular velocity curve calculated using all mass components in the galaxy. The red dashed line depicts the curve fitted with Equation (\ref{eq:Vr}). The red-shaded region illustrates the hypothetical error margin for the circular velocity (20 km $\mathrm{s^{-1}}$).\label{fig:vcirc_RC}}
\end{figure}

In section \ref{sec:data&analysis}, we analyze the rotation curves of galaxies in a way similar to observations by obtaining line-of-sight velocity maps. Here, we take advantage of using simulations to study the rotation curves of galaxies by analyzing the circular velocity curve; i.e. we exactly know the position and velocity of each particle including the total mass of galaxies.

The circular velocity of a particle rotating around the enclosed mass in a circular orbit is given by the equation: 

\begin{equation}
\label{Vcirc}
    V_\mathrm{circ}=\sqrt{\frac{GM(<r)}{r}}
\label{eq:Vcirc}
\end{equation}

where G is universal gravitational constant, r is the distance between the test particle and the center of the galaxy, and M($<r$) is the enclosed mass at the distance r. We consider the spatial position of the particle with the minimum gravitational potential energy as the center of the galaxy. This circular velocity differs from the velocities obtained by observations, as it represents the velocity obtained under the assumption that objects moving around the mass are entirely rotation-supported. The circular velocity represents the mass distribution directly, while the observed velocity is also affected by the velocity dispersion of the galaxy as shown in Equation (\ref{eq:Vdisp}). 

We calculate the circular velocity at various radii up to 3 times the stellar half-mass radius ($3\times r_{\text{half}}$) using Equation (\ref{eq:Vcirc}) with the  position and mass data of the particles within the galaxy. We fit these circular velocity curves using Equation (\ref{eq:Vr}) and obtain the parameters describing the shape of the rotation curve including $s_\mathrm{out}$. We use the same sample of 3066 galaxies for the results with the rotation velocity, $V_\mathrm{rot}$. 

Figure \ref{fig:vcirc_RC} shows an example (SubhaloID 125031) of the rotation curve derived with the circular velocity. Each mass component is converted to the velocity term with Equation (\ref{eq:Vcirc}) and is depicted in colored lines. The black squares represent the sampled circular velocities with a bin size of 1.36 kpc. We fit the sampled circular velocity with Equation (\ref{eq:Vr}) to obtain the parameters related to the shape of the curve. The circular velocity profile with the black solid line and the fitted profile with the red dashed line are in good agreement.

\begin{figure}[t!]
\centering
\includegraphics[width=80mm,scale=1]{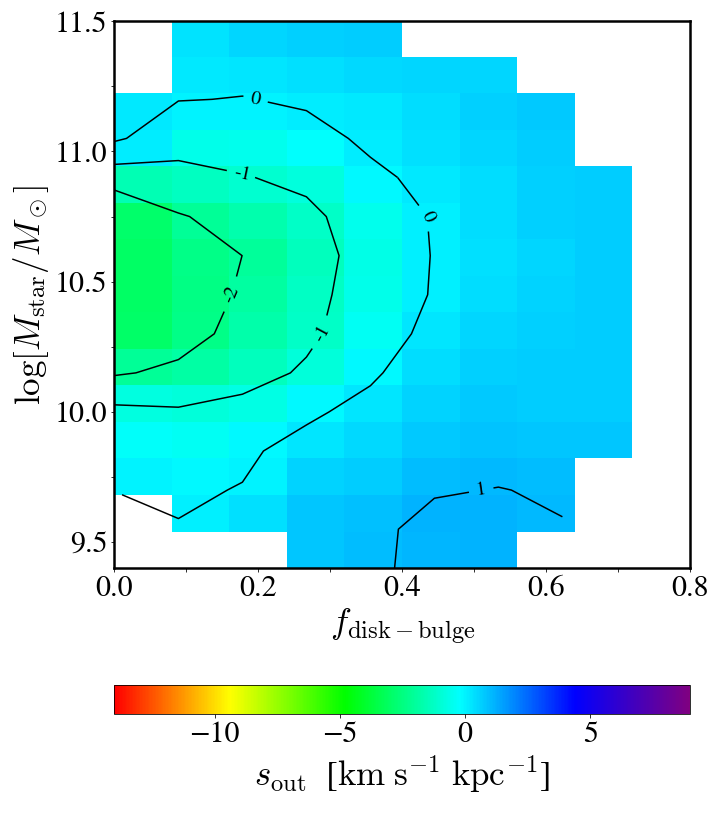}
\caption{The outer slopes of rotation curves ($s_\mathrm{out}$) fitted from circular velocity ($V_\mathrm{circ}$) in morphology versus log $M_\mathrm{star}$ plane. The outer slopes ($s_\mathrm{out}$) are represented by colors (see the color bars for the color-coded $s_\mathrm{out}$) and contours (values in the middle of contour lines indicate $s_\mathrm{out}$). Other descriptions are identical to
those in Figure \ref{fig:mainresult}. \label{fig:vcirc_sout}}
\end{figure}

We use the results from this fit to show again the dependence of outer slope on galaxy morphology and stellar mass in Figure \ref{fig:vcirc_sout}. The overall shape looks similar to Figure \ref{fig:mainresult}, but the change of $s_\mathrm{out}$ appears more mild; this could be because of the following reasons.

First, the velocity dispersion could affect the different values of  $s_\mathrm{out}$. The velocity dispersion in the disk can induce the pressure directed outward, thereby diminishing the centripetal force and, subsequently, the rotational velocity. Considering a constant velocity dispersion ($\sigma_0$) within the exponential disk under hydrostatic equilibrium, the observed rotation velocity (\citealp{Burkert_2010,Lang_2017,genzel2020rotation,yoon2021rotation}) can be given by

\begin{equation}
\label{Vobs_dispersion}
    V^2_\mathrm{obs}(r)=V^2_\mathrm{circ}(r) - 2\sigma_0^2 \left (\frac{r}{R_d}\right).
\label{eq:Vdisp}
\end{equation}
The exponential scale length, denoted as $R_d$, is related to the effective radius ($R_e$) by the equation $R_e = 1.68\times R_d$ (\citealp{Burkert_2010, yoon2021rotation}). Referring to Equation (\ref{eq:Vdisp}), in the case of a constant velocity dispersion, the impact of the velocity dispersion term on the observed velocity becomes more significant as the radius increases from the center of the galaxy. The decrease in velocity may lead to a steeper decline in the observed rotation curve, consequently yielding a smaller value for $s_\mathrm{out}$ in dispersion-dominated galaxies. This can explain why $s_\mathrm{out}$ derived from the mock velocity maps tends to be lower than that from the circular velocity for the galaxies affected by the significant amount of velocity dispersion (i.e. low $f_\mathrm{disk-bulge}$).

Furthermore, the fitted boundary in radius could impact on the discrepancy in the slope of the outer region of the rotation curve. In the case of the mock observed velocity, we use only the stellar particles and set the signal-to-noise cut to obtain reliable velocity map. Due to the mass resolution limit of the simulation, there are few stellar particles at the outer region of galaxies for smaller and less massive galaxies. As a result, the velocity map of these galaxies might not have a high enough number of data points out to 3 times the stellar half-mass radius ($3\times r_\mathrm{half}$), resulting in the fit without the data in the outer part of galaxies. 
On the other hand, we use all the particles in a galaxy without applying any signal-to-noise cut when deriving circular velocity curves. Hence, the circular velocity curves of all galaxies are fitted up to 3 times the stellar half-mass radius. This difference in the fitted radius could affect the dissimilarity between the fitted $s_\mathrm{out}$ values of the mock observed velocity curve and the circular velocity curve. \\

\subsection{Dark Matter Density Profile} \label{subsec:DMdensity}
\begin{figure}[b!]
\centering
\includegraphics[width=80mm,scale=1]{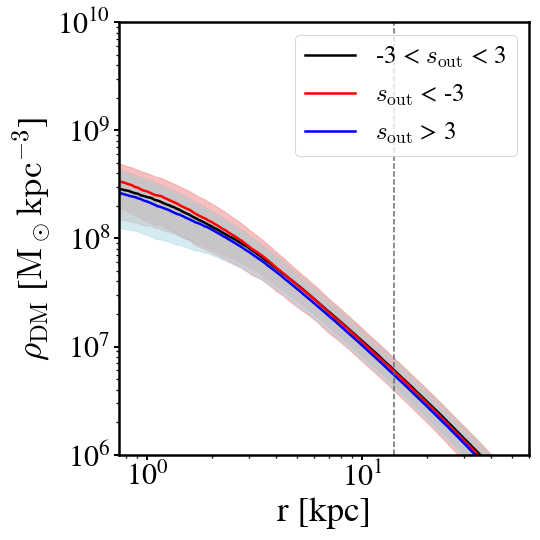}
\caption{Dark matter density profiles of galaxies in a specific mass bin (11.5$<$ log[$M_\mathrm{DM}/M_\odot$] $<$12.0). Colors denotes the value of $s_\mathrm{out}$ of the galaxies. The shaded region show the standard deviation of the dark matter density profiles of the galaxies. The grey dashed line shows three times the median stellar half-mass radius of the final sample of $V_\mathrm{y}$ line-of-sight velocity. \label{fig:dmdensity}}
\end{figure}

\begin{figure*}[hbt!]
\centering
\includegraphics[width=160mm,scale=1]{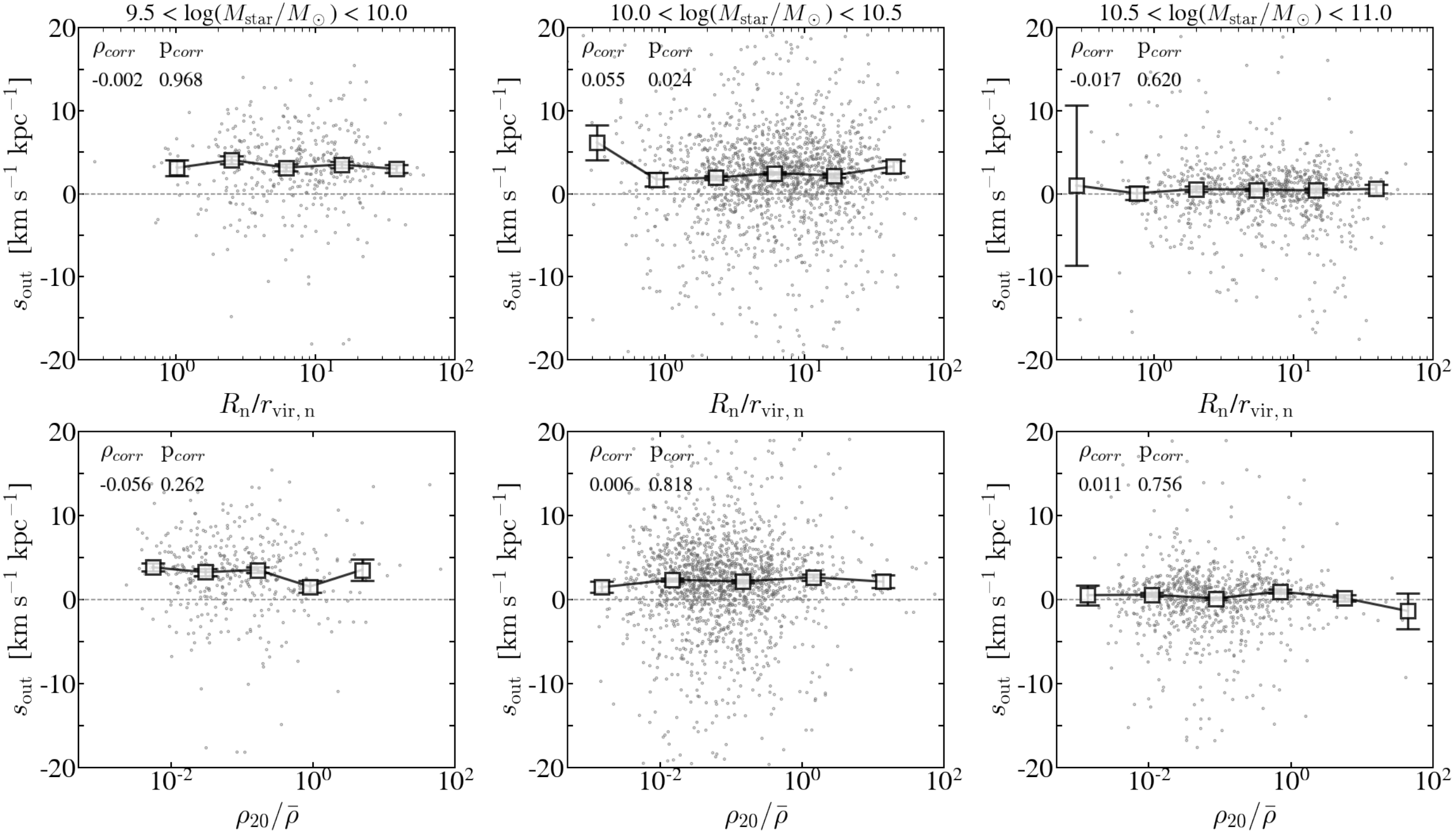}
\caption{Dependence of the slope of the outer region of rotation curves ($s_\mathrm{out}$) on the environmental parameters. The top panels are for the pair separation between target galaxies and their nearest neighbors. The bottom panels are for the large-scale background density. The target galaxies in the
left, middle, and right panels are limited to their stellar mass ranges of log ($M_\mathrm{star}/M_\odot$) = 9.5–10.0, 10.0–10.5, and 10.5–11.0, respectively. The error bar is the standard deviation of the median values from 1000 bootstrap resamplings. The numbers in the top left-hand corner of each panel denote the Spearman rank correlation coefficient and the probability of obtaining the correlation by chance. \label{fig:env_sout}}
\end{figure*}
The Navarro-Frenk-White (NFW) profile (\citealp{NFW96}) built upon simulations was found to be a good fit for a wide range of halo masses in numerical simulations within the context of the Cold Dark Matter (CDM) cosmological model. However, several studies have emphasized limitations of it, especially for the dark matter profiles of subhaloes, and delved deeper into the exploration of dark matter profiles. 

\citet{DiCintio13} suggest that the Einasto model (\citealp{Einasto65}) offers a superior description of the density profile of subhaloes compared to the more commonly employed NFW profile. They also demonstrate a strong correlation between the shape parameter of the Einasto profile and the total mass of the subhalo. Some studies show that the mass and the tidal stripping affect the evolution of the dark matter density profile in galaxies (\citealp{Green19,DiCintio13}). To assess the influence of distinct shapes in the dark matter density profile on diverse shapes of rotation curves, we analyze dark matter density profiles across different values of the outer slope of the rotation curve ($s_\mathrm{out}$).

Figure \ref{fig:dmdensity} shows the median dark matter density profiles for three samples with different $s_\mathrm{out}$ values. We fix the range of dark matter mass to remove any mass effect on density profile and on $s_\mathrm{out}$: 11.5$<$ log[$M_\mathrm{DM}/M_\odot$] $<$12.0 where the number of galaxies in a bin is the largest (i.e. 1401 galaxies). The colors denote the values of $s_\mathrm{out}$. The dark matter density profiles for the samples with different $s_\mathrm{out}$ are indistinguishable within the standard deviations. Although the three different profiles are almost identical at the outer part, they show a small difference at the inner part of the galaxies. The red profile that denotes the declining rotation curve has the highest value of the dark matter density in the inner region. The blue profile that represents the ascending rotation curve has the lowest value of dark matter profile in the inner region. This result appears consistent with the expectation that a galaxy exhibiting a decreasing (increasing) rotation curve would contain more (less) concentrated matter in its central region. Nevertheless, our findings indicate that the shape of the dark matter density profiles does not determine the outer slope of galaxy rotation curves, given the considerable standard deviations of the dark matter density profiles.

\subsection{Environmental dependence of $s_\mathrm{out}$} \label{subsec:environment}

In this section, we would like to examine whether the shape of galaxy rotation curves is also dependent on environment in addition to their internal properties. To have a fair comparison with observational data and to characterize the full range of galaxy environments, we use two different environmental parameters: $R_\mathrm{n}/r_\mathrm{vir,n}$ and $\rho_{20}/\bar{\rho}$ (\citealp{Park08,Muldrew12}). 

$R_\mathrm{n}/r_\mathrm{vir,n}$ is the distance to the nearest neighbor galaxy normalized with the virial radius of the neighbor galaxy, which is a relatively small-scale environmental parameter. We define the nearest neighbor galaxy of a target galaxy as the one that is closest to the target and satisfies a mass condition, $M_\mathrm{neighbor}>10^{-0.2}\times M_\mathrm{target}$. This mass condition is adopted to be similar to the one used for observations based on magnitude; e.g. \citealp{Park08} identified the nearest neighbor galaxy among those brighter than the target by $M_\mathrm{r, target} + 0.5$. This condition is introduced to consider only the galaxies that can exert gravitational influence upon the target galaxy. The virial radius of a galaxy is calculated as the radius of the sphere where the mean mass density becomes 200 times the critical density (\citealp{Park09}).

Another parameter is $\rho_{20}/\bar{\rho}$ that is a normalized background mass density, which is a relatively large-scale environmental parameter. The background density at a galaxy location is measured by 

\begin{equation}
    \rho_{20}/\bar{\rho}=\sum\limits_{i=0}^{20}M_i W_i(|r_i|, h)/\bar{\rho} 
\label{eq:rho20}
\end{equation}

using the total mass $M$ of the 20 closest galaxies in the simulation. The measurement for the galaxies located at the boundary of the simulation box is conducted correctly with the simulation box volume enlarged using a periodic boundary condition. The weighting $W_i$ is the standard kernel of smooth particle hydrodynamics (SPH) for density calculation (\citealp{Monaghan85, Muldrew12}). The mean mass density within the simulation volume V is measured by

\begin{equation}
    \bar{\rho}=\sum\limits_{all}M_i /V
\label{eq:meanrho}
\end{equation}

where the summation is over all galaxies in the simulation.

Figure \ref{fig:env_sout} displays the relation between the outer slope of the rotation curves and the environments of our sample galaxies. To remove the mass effect, we make three subsamples using narrow ranges of galaxy mass (i.e. three columns). It is hard to find a meaningful correlation between the outer slope of the rotation curves and the environments of the galaxies. The spearman correlation coefficients are close to zero and their p-value are also quite large. These results suggest that the galaxy environments do not have significant impact on the shape of galaxy rotation curves.

\section{conclusions} \label{sec:summary}

We use the IllustrisTNG cosmological hydrodynamical simulation to study the diverse rotation curves of galaxies at z=0. We conduct the fit to determine the rotation velocity from the two-dimensional velocity map that we have constructed. We successfully reproduce the observed dependence of rotation curve shape on the stellar mass and the morphology. Our main results are as follows.
\begin{enumerate}
    \item[(i)] The outer slope ($s_\mathrm{out}$) of the rotation curve increases as galaxies are more disky, and decreases as galaxies are more massive, except for the very massive early-type galaxies ($M_{*}> 10^{10.5}M_\odot$). Early-type galaxies with a stellar mass around $10^{10.5}M_\odot$ exhibit the smallest $s_\mathrm{out}$ value.
    \item[(ii)] We find a hint that the dark matter fraction is more important in determining the outer slope of galaxy rotation curves ($s_\mathrm{out}$) than the gas fraction. 
    \item[(iii)] The overall trend of $s_\mathrm{out}$ from the circular velocity curve is similar to that from the mock observed velocity curve; but the change of $s_\mathrm{out}$ is more mild. The difference could result from the effects of the velocity dispersion and the radius limit for the fit.
    \item[(iv)] The dark matter density profiles of the galaxies with different $s_\mathrm{out}$ show no noticeable difference across the galactocentric radius.
    \item[(v)] The correlation between galactic environments and the outer slope of galaxy rotation curves is not statistically significant.
\end{enumerate}
These results suggest that our simulation samples can serve as an important testbed for the subsequent study tracing galaxies back in time, enabling a deeper understanding of the physical origin behind the diverse rotation curves. In this regard, we plan to trace the mass accretion history of each galaxy to better understand how the galaxy rotation curves reach their current shape. \\




\section*{Acknowledgments}
We thank the referee for insightful comments that improve the paper.
We also thank the IllustrisTNG collaboration for making their simulation data publicly available. 
H.S.H acknowledges the support of Samsung Electronic Co., Ltd. (Project Number IO220811-01945-01), the National Research Foundation of Korea (NRF) grant funded by the Korea government (MSIT), NRF-2021R1A2C1094577, and Hyunsong Educational \& Cultural Foundation.

\bibliography{refs}{}

\bibliographystyle{aasjournal}



\end{document}